\documentclass[letterpaper,twocolumn,10pt]{article}
\usepackage{usenix-2020-09}

\usepackage{shortcuts}
\usepackage{xspace}
\usepackage{lipsum}
\usepackage{multirow}
\usepackage{colortbl}
\usepackage{tabularx}
\usepackage{subcaption}
\usepackage{colortbl}
\usepackage{subcaption}
\usepackage{booktabs}
\usepackage{graphicx}
\usepackage[usestackEOL]{stackengine}
\usepackage{stfloats}
\usepackage{balance}
\usepackage{url}
\usepackage{soul}

\renewcommand{\paragraph}{\vspace{3pt}\noindent\textbf}
\begin{document}

%
%

\title{One Size Does not Fit All: Quantifying the Risk of Malicious App Encounters for Different Android User Profiles}

\author{ Savino Dambra$^*$, Leyla Bilge$^*$, Platon Kotzias$^*$, Yun Shen$^\dag$, Juan Caballero$^\ddag$\\ 
	$^*$Norton Research Group \quad $^\dag$NetApp \quad $^\ddag$IMDEA Software Institute\\
	}
\maketitle

\begin{abstract} \noindent 

Previous work has investigated the particularities of security practices within
	specific user communities defined based on country of origin, age, prior
	tech abuse, and economic status.  Their results highlight that current
	security solutions that adopt a one-size-fits-all-users approach ignore the
	differences and needs of particular user communities.  However, those works
	focus on a single community or cluster users into hard-to-interpret
	sub-populations.
In this work, we perform a large-scale quantitative analysis of the risk of
	encountering malware and other potentially unwanted applications (PUA)
	across user communities.  At the core of our study is a dataset of app
	installation logs collected from 12M Android mobile devices.
Leveraging user-installed apps, we define intuitive profiles based on users'
	interests (e.g., gamers and investors), and fit a subset of 5.4M devices to
	those profiles.
Our analysis is structured in three parts.  First, we perform risk analysis on
	the whole population to measure how the risk of malicious app encounters is
	affected by different factors.
Next, we create different profiles to investigate whether risk differences
	across users may be due to their interests. 
Finally, we compare a per-profile approach for classifying clean and infected
	devices with the classical approach that considers the whole population.
We observe that features such as the diversity of the app signers and the use of
	alternative markets highly correlate with the risk of malicious app
	encounters.  We also discover that some profiles such as gamers and
	social-media users are exposed to more than twice the risks experienced by
	the average users. We also show that the classification outcome has a marked
	accuracy improvement when using a per-profile approach to train the
	prediction models.  
Overall, our results confirm the inadequacy of one-size-fits-all protection
	solutions.
\end{abstract}

\section{Introduction} \label{sec:intro}
In recent years, researchers conducted a large number of user studies to
investigate the particularities of security and privacy practices within
specific user communities defined based on their
country of origin~\cite{10.5555/3235924.3235943,10.5555/3489212.3489322},
age~\cite{238333,10.5555/3235924.3235943,10.1145/2858036.2858317},
disabilities~\cite{244030}, professions~\cite{thomas21usenix,190976}, economical
status~\cite{48171,10.1145/3274445}, abuse-survivor
condition~\cite{235451,10.1145/3359304}, and political
exposure~\cite{10.1145/3173574.3173688}.  One of the most common takeaways of
these studies is that current security-, privacy-, and digital-safety solutions
fail to consider classes of at-risk users, 
who, due to their unique user characteristics,
might encounter cyber threats more frequently or 
may encounter different cyber threats (e.g., more
sophisticated)~\cite{warford2022sok}.

Prior work discusses the problems of the current one-size-fits-all approach and
how the digital needs of particular populations receive less attention from the
security community~\cite{onesizefitsall}.  Following these invaluable insights,
a few recent works target a specific
community~\cite{8418618,10.5555/3489212.3489320,10.1145/3313831.3376747,
kevincreepware}.  However, these works should be supported by large-scale
studies that can quantify and compare cyber-risk levels across different
communities.  Results from such studies could motivate \vendors to provide more
personalized security and privacy solutions to help particular at-risk
communities. 
Such large-scale quantitative analysis for customized security and risk analysis
has been explored in enterprises~\cite{olivierrisk}.  However, enterprises are
restricted environments with their own risk factors such as enterprise size,
geographical location, assets owned, and the industrial sector they belong
to~\cite{enterprise}. 
Unfortunately, those indicators do not directly translate to consumers. 
Furthermore, the set of computers that belong to an
enterprise is well-defined.  In contrast, large-scale identification of
individual users that belong to the same community (e.g., users with shared
interests), without a priori user knowledge, is a challenging problem.  Thus,
researchers restricted their analysis to small sets of users that are interviewed
about who they are, what they do, and what they like. 

In this work, we perform a large-scale quantitative analysis of the risk of
encountering malware and other potentially unwanted applications (PUA) 
across user communities. 
We group users in the same community if they share interests.
Communities are identified based on information collected in a 
privacy-sensitive manner from their mobile devices. 
Mobile devices have some unique characteristics that
make them a great source of data to identify user communities. 
First of all, mobile devices are most often personal, 
although there are known exceptions such as 
device sharing within households~\cite{bodker2012poetry,matthews2016she} and
in certain geographical locations such as
South Asia~\cite{ahmed2013ecologies,sambasivan2018privacy,ahmed2019everyone,al2021we}
and Africa~\cite{murphy2011my}.
Being portable, they are
typically kept close to their users throughout their daily routine and have become an
indispensable accessory for many activities such as getting informed, listening
to music, consuming video content, creating art, booking travel, and ordering
food.  Furthermore, mobile devices are ubiquitous being used by 3.8B people,
with an average user spending 5.4 hours a day using them, 
and 13\% of the millennials over 12 hours~\cite{techjury}.

In the smartphone world, users install a wealth of mobile applications (apps for
short) on their mobile devices to enable the above activities.  As such,
the apps installed on a mobile device characterize its usage 
and thus can reveal user interests, needs, and personality traits,
e.g., better than the set of programs installed in a desktop
computer~\cite{zhao2019user}.  Motivated by this, recent studies leverage app
installation logs to profile mobile users and infer users' gender or
age~\cite{ying2012demographic, zhao2019gender, seneviratne2014predicting,
seneviratne2015your}, income~\cite{zhao2019user}, and
interests~\cite{zhao2016mining}.
Similarly, at the core of our study is a dataset of app installation logs
collected in a privacy-sensitive manner by a \vendor from 12M Android mobile
devices (8.6M after filtering).

Our main goal is to shed light on risk factors correlated with the encounter of
malware or PUA, and assess the risk faced with respect to those factors by
different user profiles. While risk assessment represents one of the
cornerstones of computer security, very few studies provide quantitative
findings by leveraging large and real-world empirical data. Our mid-term goal,
on the other hand, is to promote more personalized security and to foster
personalized risk assessments that put each user and their specific needs at the
center by abandoning previous models and assumptions that consider the
whole population as a unique entity to protect.  

We start by categorizing Android users based on the apps they install 
into intuitive user profiles. 
Identifying user communities in the app installation logs 
is a challenging problem as the logs do not contain any information 
about the users that own the devices.
Prior work has inferred user interests from app installations by applying
unsupervised machine learning (ML) to cluster devices with similar apps
installed~\cite{seneviratne2015your,zhao2016mining,seneviratne2014predicting}.
Unfortunately, due to the ML black-box nature,
it is not clear what user profile each cluster represents.
In contrast, we define intuitive user profiles such as gamers and
investors, and look for devices that fit those profiles.
For this, we assume that users that install
multiple apps from some selected categories
(e.g., \texttt{Games}, \texttt{Finance}),
have a shared interest and thus a common profile (e.g., gamer, investor).
We also build profiles for users with more than one interest.
For comparison, we include two additional profiles: 
average users and users with mixed profiles 
(i.e., with strong interests in more than two categories).
The generated profiles cover 5.4M devices.

Our analysis is structured in three parts. 
First, we perform single-feature risk analysis on the whole population 
of 8.6M filtered users to measure how the risk of malicious 
app encounters is affected by 10 risk factors.
As much as our data allows, we study similar features of previous studies 
that focused on the risk estimation of desktop computers and their
users~\cite{riskteller,canali2014effectiveness,nappa2015clones,sharif2018predicting,andropup}.
To the best of our knowledge, 
our work is the first to explore this topic on the mobile ecosystem.
Next, we investigate whether risk differences across users may be
due to their interests, i.e., to behaviors that users with common
interests may share.
For this, we perform 
a profile-based risk analysis
to assess whether each profile presents cyber risks that are considerably
different than those of the average users, who are the main target
of current security solutions.
Finally, we perform multi-feature risk analysis 
for investigating the contribution of risk factors
when classifying clean and infected devices. 
We observe significant improvements in the classification accuracies
when employing dedicated profile models to predict the risk scores of mobile devices.
Such risk scores are fundamental for personalized security,
 privacy, and online safety solutions. 
For example, they can be used to adjust the AV engine settings 
(e.g., scan frequency) and to improve security notifications~\cite{harbach2014using,liu2016follow,wisniewski2017making}.

This paper provides the following main insights:

\begin{itemize} 
\item The diversity of the application
			signers and the use of alternative markets highly correlate with the
			risk of detecting malicious samples. While less marked, 
			app reputation and update frequency show a not negligible
			contribution on the total risk.  
\item Users assigned to specific
				profiles show significant differences when compared to the
				general population. While some maintain a lower infection rate, 
				others such as
			\texttt{Gamers} are more impacted. Profiles also show different
		threat sources when measuring the most contributing categories of
	applications that are marked as malicious.  
\item The classification results
		reflect the inadequacy of one-fits-all protection solutions. 
		Using a single model, the accuracy score for some profiles is even
		worse than 50\%, while for others it reaches upwards of 80\%. The
		outcome of the per-profile classification approach is significantly
		better with an average accuracy of more than 76\%.  \end{itemize}


\section{Datasets} \label{sec:datasets}

This section presents the datasets that constitute the basis of our study.  At
the core of our study lies a dataset of \textit{App installation logs} from a
\vendor with metadata of apps installed on 12M Android devices.  For apps
distributed through the official \Pl, we also obtain their market metadata.
We use the \Pl category to classify the apps.  We collect VirusTotal (VT)
reports of the APKs and use the AV detection labels in the VT reports for
identifying malware, PUA, and for classifying those into families. 

\begin{table}[t]
\small
\centering
\caption{Summary of datasets used.}
\label{tab:datasets}
\begin{tabular}{llrr}
\hline
\textbf{Dataset} & \textbf{Data} & \textbf{All} & \textbf{Filtered} \\
\hline
App Installation Logs          & Devices                 &  12.2~M & 8.6~M \\
                               & Countries               &  243    &  243 \\
\cline{2-4}
                               & APKs                    &  34.6~M & 8.9~M \\ 
                               & Packages                &   7.9~M & 2.2~M \\ 
															 & Signers                 &   4.1~M & 1.5~M \\ 
\hline
VirusTotal                     & Reports                 & 4.8~M   & 875~K \\ 
\hline
\end{tabular}
\end{table}






\paragraph{App installation logs.} These logs capture metadata about the
presence of apps in 12M Android devices.  These logs are collected from real
devices in use by customers of the \vendor that opted-in to share their data and
are anonymized to preserve customer privacy.  We discuss ethical concerns in
Section~\ref{sec:discussion}.  Each device in the dataset regularly queries a
backend system to obtain the reputation of the APKs installed in the device.
The dataset includes only app metadata and not the actual APK files.  The
dataset covers four months from June 1st, 2019 to September 30, 2019.
It consists of file metadata such as APK hash, APK package name, the
signer (i.e., the SHA256 of the public key in the APK's certificate), and the
name of the parent package that installed the APK.  The parent information is
obtained via Android's \emph{PackageInstaller} class using the
\textit{PackageInstaller.getInstallerPackageName} method.  Parent information
may not be available for all apps.  For example, apps that come pre-installed on
the device or sideloaded apps installed via the Android Debug Bridge (ADB) may
not have parent information.  The dataset also contains a list of countries from
where a device has connected to the backend service, obtained through
geo-location of the device's IP address.
An APK may have been queried by the same device more than once.  We obtain the
earliest date that an APK is observed in a device and use it as an approximation
of the installation time for that app version. 
As shown in Table~\ref{tab:datasets}, the
dataset contains 34.6M APKs that belong to 7.9M packages collected from 12.2M
Android devices in 243 countries.

\paragraph{\Pl.} For app categorization, we use \Pl metadata, namely the app's
category.  The \Pl metadata was collected in February 2020
by trying to download the app's public webpage using its package name. 
Of the 7.9M packages in the app installation logs, 24\% (1.9M) were present in
\play. This fraction is similar to that measured in 
previous works~\cite{preinstalled,andropup}.
Apps not in \Play include 
AOSP packages, 
apps pre-installed by the vendor or carrier, 
apps available in the past but since removed~\cite{wang2018android}, and 
apps distributed through alternative vectors 
(e.g., other markets, browser downloads).
After filtering apps not installed by the user (see Section~\ref{sec:approach})
the percentage of categorized apps on the filtered dataset raises to 53\%.

\paragraph{VirusTotal.} 
We query the hash of APKs in VirusTotal (VT)~\cite{vt}.
VT offers a commercial API that given a file hash
returns file metadata and the list of detection labels assigned by a large
number of AV engines used to scan the file.  
Unfortunately, given VT's API restrictions, we could not query all APKs.
To minimize any bias and to increase our coverage 
we query the ten most prevalent APKs for each app signer and all APKs 
that install other apps.
In total, we collect VT reports for 14\% (4.8M) of all the APKs and 
31\% (2.5M) of all apps in the dataset.
However, after the dataset filtering is done (see Section~\ref{sec:approach}),
we remain with VT reports for 10\% (875K) of all APKs and 20\% (443K) of all apps.
On a per-device level, we measure that for 35\% of the devices we have obtained
VT reports for at least 20\% of the installed apps.
We present the whole distribution of the per-device VT coverage 
in Figure~\ref{sec:appendix-fig:vt_dev_distr} in the Appendix.
We use the AV labels from the VT reports to identify malware and PUA.

\section{Dataset Preparation} \label{sec:approach}

\paragraph{Device and app filtering.} Before performing the risk analysis, we
apply a number of filtering steps on the app installation logs and then identify
malware and PUA among the remaining data.  At a high level, the filtering
comprises four steps and has two goals.  First, three filters are applied for
removing devices that do not capture real user behavior.  These include devices
used for tasks such as app testing or research, and devices for which there is
not enough information to capture the behavior of its user (e.g., devices with
very few apps).  As done by previous work~\cite{kevincreepware}, we empirically
selected these thresholds by analyzing the distribution of installed apps per
device and looking for outliers for varying outlier cutoff values.  Then, a
fourth filter aims at removing apps that have not been installed by the user
(e.g., OS apps, pre-installed apps) and thus cannot be used to characterize the
user.  We outline their technical details below.

\begin{itemize}
	
\item The first filter removes devices with an excessive number of installed
	apps since they are most probably devices used for testing purposes.  More
		specifically, we filter out 60K devices that have more than 500 apps
		installed. 

\item The second filter removes devices that query the backend system only once
	within our analysis period.  These devices likely uninstalled the AV app
		promptly after installation.  No full scan was likely performed on those
		devices and therefore there is limited visibility on their list of apps.
		This step removes 1.9M devices.

\item The third filter removes devices with a very small number of installed
	apps, e.g., devices that turned off the telemetry consent.  This step
		removes 1.4M devices with less than 15 installed apps including the OS
		and other system apps.

\item The fourth filter removes apps whose installation is not rooted on a
	decision by the device user. 

In Section~\ref{sec:behavioral}, we perform a systemic risk analysis to
		investigate whether certain user profiles diverge from the risk models
		made for the whole population.  For this, we group users based on the
		kinds of apps they install on their devices.  However, not every app on
		a device has been installed by its user.  Some apps are already present
		when the device is acquired such as those belonging to the Android Open
		Source Project (AOSP) and those pre-installed by the device manufacturer
		and the carrier~\cite{preinstalled}.  Other apps may be downloaded
		without user intervention by system-level apps, e.g., as part of
		firmware upgrades~\cite{blazquez2021trouble}. 
Since the user did not originate the installation of those apps, their presence
		does not provide information about user intentions.  Thus, they should
		not be included in the user's categorization.
However, identifying user-installed apps is challenging.  OS restrictions
		prohibit apps from monitoring such events.  System-level apps are
		installed under a system partition (e.g., \textit{/system},
		\textit{/vendor}, \textit{/oem}), but the installation path is not
		available in our dataset.  To address this issue, we consider as
		user-installed apps those installed from the official or alternative
		markets.  The intuition is that AOSP and system-level apps are not
		distributed through markets, but come pre-installed and are updated
		through firmware upgrades~\cite{blazquez2021trouble}.  We may miss
		user-installed apps distributed through other vectors such as IM and
		browser downloads.
But, according to a prior study, 93\% of app installations come from
		markets~\cite{andropup}.  Thus, this filter will maintain a good
		representative list of user-installed apps for each device.  To identify
		market-originated apps, we match the parent package against a list of
		122 market packages (and their signers) produced in prior
		work~\cite{andropup}.  We detail limitations of our user-installed app
		detection in Section~\ref{sec:discussion}.

\end{itemize}

\noindent As presented in Table~\ref{tab:datasets}, the final filtered dataset
contains 8.9M APKs from 2.2M packages found on 8.6M devices.

\paragraph{Malware and PUA detection.} To measure the risk of malware and PUA
installations, we first need to identify them in the dataset. 
A common practice is to collect AV detection labels from VT and consider
malicious any sample flagged (i.e., assigned a non-empty label) by at least a
threshold number of AV engines.  A higher threshold reduces false positives due
to a few AV engines making an incorrect determination, but may increase false
negatives.  Recent work has shown that threshold values between 2 and 14 are
good for stability and for balancing precision and
recall~\cite{zhu2020measuring}.  We use a threshold of at least 4 AV engines,
which is within the recommended range and has been used by a variety of other
works~\cite{malsign,ppipup,andropup}.
From the 4.8M APKs for which a VT report was collected, 28.8\% (1.5M) have zero
detections, 12.8\% (681.4K)  between one and three detections,
and 58.5\% (3.1M) have at least four detections 
will be the focus of the risk analysis. 
To distinguish between malware and PUA, we feed the VT reports of the 3.1M
detected apps to the AVClass malware labeling tool~\cite{avclass}.  AVClass
outputs the most likely family name 
classifies the sample as PUA or malware based on the presence of PUA-related
keywords (e.g., grayware, adware, pua, pup).
AVClass classifies 73.4\% of the 3.1M APKs as PUA and the remaining 
26.6\% as malware.

\section{Android User Risk Factors} \label{sec:systematic} 
This section details our
global risk analysis on the whole population for identifying risk
indicators and their contribution.

To assess the cyber risk of users, previous works have investigated a wide range
of
features~\cite{riskteller,canali2014effectiveness,nappa2015clones,sharif2018predicting,andropup}.
Those features can be grouped into five classes: features that measure the
volume of online activity (e.g., number of websites browsed or files
downloaded); those reflecting the diversity of a user's online behaviors (e.g.,
category of websites browsed and programs downloaded); those capturing the
reputation of entities the user interacts with (e.g., program publishers,
browsed websites); those capturing software updates (e.g., vulnerability
patching rate); and geographical features (e.g., user country). 
In this section, we test to what extent similar risk trends affect Android users.
We focus on feature classes because we cannot use exactly the same features
evaluated in prior work since our platform, dataset, and analysis time period
are considerably different. 

We measure how 10 features, representative of the above 5 feature classes,
correlate with malicious encounters in the Android ecosystem.  For this, we
build three Generalized Linear Models (GLMs)~\cite{cameron2013regression}, each
of them modeling 10 independent variables (one per feature) by using a binomial
distribution and a logarithmic link function.
Each model captures respectively the outcome of detecting (1) any malicious app
(i.e., malware or PUA), (2) only malware, and (3) only PUA. 
Once fitted to the data, each model outputs a Log-Odds distribution for the
dependent variable $Y$ that expresses the odds
of a malicious encounter as a linear combination of the 10 features.  The extent
to which each feature influences a malicious encounter is captured by its
regression coefficient.  We reduce granularity by bucketing each feature into
quartiles (i.e., 4 bins).
We set the first bin as reference, a common choice in related
works~\cite{simoiu2020targeted, van2020go}, and express the odds ratios of other
bins with respect to the reference.

To select the best model, we test different GLM configurations and analyze the
outcome of several goodness-of-fit quantities (Pseudo R-Squared, Log-Likelihood,
Dispersion) in addition to controlling for feature dimensionality by employing the
Akaike Information Criterion (AIC). 
Table~\ref{sec:systematic-tbl:glm} summarizes the results and we discuss each
feature class in the following paragraphs. In addition,
in section~\ref{appendix:best_worst} of the Appendix, we conduct an in-depth
analysis to evaluate how the identified factors differ between the best and worst devices in our
dataset (i.e., clean devices exhibiting no signs of malicious apps for all the
period of our experiments, and outlier devices that report a very high number of
malicious detections)

\begin{table*}[!th]
	\small
	\centering
	\setlength\extrarowheight{-2.5pt}
	\caption{Increase in odds of encountering malware, PUA or any form of
	malicious application (PUA \& malware) according to our binomial regression model. All values
	have significance $p < 0.0001$}
	\label{sec:systematic-tbl:glm}
	\begin{tabularx}{0.9\textwidth}{l|l|cc|ccc}
\toprule
		\multirow{2}{*}{\textbf{Feature}}&
		\multirow{2}{*}{\textbf{Class}}&
		\multirow{2}{*}{\textbf{Bin}}&
		\multirow{2}{*}{\textbf{Reference}}&
		\multicolumn{3}{c}{\textbf{Odds ratios}} \\
		\multicolumn{4}{c}{} &
		\multicolumn{1}{c}{\textbf{PUA}}&
		\multicolumn{1}{c}{\textbf{Malware}}&
		\multicolumn{1}{c}{\textbf{PUA  \& Malware}}\\
\midrule
\multirow{3}{*}{Applications} & \multirow{3}{*}{Volume}					&          31-52 &        	15-30 &      1.24 &          1.15 &      1.24 \\
			                           			& &         53-102 &        	15-30 &      1.31 &          1.28 &      1.33 \\
			                           			& &           >102 &        	15-30 &      1.36 &          1.62 &      1.38 \\
\midrule
\multirow{3}{*}{Activity days}  & \multirow{3}{*}{Volume}                &           6-20 &         2-5 &      1.13 &          1.07 &      1.14 \\
& 												&          21-49 &         2-5 &      1.36 &          1.30 &      1.37 \\
& 												&            >49 &         2-5 &      1.38 &          1.51 &      1.41 \\
\midrule
\multirow{3}{*}{Signers}  & \multirow{3}{*}{Diversity}                      	&          26-50 &        	1-25 &      1.89 &          1.80 &      1.91 \\
& 												&          51-75 &        	1-25 &      3.94 &          3.29 &      4.00 \\
& 												&            >75 &        	1-25 &      9.22 &          7.03 &      9.42 \\
\midrule
\multirow{3}{*}{Application categories} & \multirow{3}{*}{Diversity}		&          13-16 &        1-12 &      0.77 &          0.69 &      0.75 \\
& 								                &          17-20 &        1-12 &      0.52 &          0.48 &      0.50 \\
& 								                &            >20 &        1-12 &      0.40 &          0.38 &      0.38 \\
\midrule
\multirow{3}{*}{Update rate}  & \multirow{3}{*}{Updates}                  	&        Medium  &     		High &      1.19 &          1.46 &      1.22 \\
& 												&        Low     &       	High &      1.53 &          1.91 &      1.58 \\
& 												&        Very Low&  	  	High &      1.99 &          2.75 &      2.07 \\
\midrule
\multirow{3}{*}{\% Apps from Alternative markets}  & \multirow{3}{*}{Distribution}&        26\%-50\% &        	0\%-25\% &       0.92 &          0.78 &      0.90 \\
& 												&        51\%-75\% &        	0\%-25\% &       1.82 &          1.47 &      1.78 \\
& 												&           >75\% &        		0\%-25\% &      8.21 &         12.03 &      7.68 \\
\midrule
\multirow{3}{*}{Alternative markets} & \multirow{3}{*}{Distribution}			&            1   &         		0 &      1.08 &          0.97 &      1.09 \\
& 												&            2-4 &         		0 &      1.57 &          2.04 &      1.64 \\
& 												&             >4 &         		0 &     12.74 &         12.99 &     17.94 \\
\midrule
\multirow{3}{*}{App prevalence}	& \multirow{3}{*}{Reputation} &  		  Medium &    	 		Low &      0.76 &          0.79 &      0.78 \\
& 												&  			High &    	 		Low &      0.74 &          0.82 &      0.77 \\
& 												&      Very High &    	 		Low &      0.69 &          0.69 &      0.71 \\
\midrule
\multirow{3}{*}{Countries} & \multirow{3}{*}{Geography}						&            2	 &         		1 &      1.22 &          1.26 &      1.21 \\
& 												&            3-5 &         		1 &      1.47 &          1.60 &      1.44 \\
& 												&             >5 &         		1 &      2.32 &          2.40 &      2.27 \\
\midrule
\multirow{6}{*}{Continent} & \multirow{6}{*}{Geography}                     	&           Africa &     	False   &      1.54 &          1.65 &      1.58 \\
& 												&           Asia &     		False   &      0.97 &          0.92 &      0.96 \\
& 												&           Europe &     	False   &      0.62 &          0.58 &      0.62 \\
& 												&           North America & False   &      0.60 &          0.49 &      0.60 \\
& 												&           Oceania &     	False   &      0.61 &          0.41 &      0.59 \\
& 												&           South America &	False   &      1.11 &          1.33 &      1.19 \\
\bottomrule
\end{tabularx}
\end{table*}

\paragraph{Volume.} Previous work has shown that the volume of online activity
is highly correlated with the number of malicious encounters in Windows
hosts~\cite{riskteller} and in the Web~\cite{canali2014effectiveness}.  We
examine 2 volume features: the number of installed apps (i.e., package names) 
and the number of days the device is active.  We observe a similar risk increase
among Android users on both features.  The overall risk is similar to the PUA
encounter risk given the higher PUA prevalence (73\%) in the dataset.
Similar to other platforms, the more extensive use of a
device increases the attack surface and thus the risk, although the risk
increase is moderate compared to other classes. 

\paragraph{Diversity.} We examine two app diversity features: the number of app
signers and the number of app categories.  The number of app signers
represents a reliable risk indicator: users with apps belonging to 51--75
publishers have 4 times the probability of reporting malicious encounters
compared to those installing apps from less than 25 publishers.  The risk is 9
times higher when compared to users with apps from more than 75 publishers.  The
risk increase is much more pronounced for publisher diversity than for app
volume, likely because installing many apps from benign publishers does not
lower the security posture. 
Surprisingly, the number of app categories negatively correlates with malware
encounter risk.  
Previous work has shown that \play hosts a smaller fraction of
malicious apps~\cite{andropup}.  Thus, this result could be explained by the
high number of apps (47\%) that are not available in \play and thus are
uncategorized.

\paragraph{Updates.} Previous studies measured the impact of patching speed on
encountering malware on Windows hosts~\cite{riskteller}.  To examine this risk
factor in Android, we measure for each device the ratio $R$ between the number
of distinct package names  and the number of APKs installed.  A ratio of 1
indicates a single version for each app, i.e., a low update rate.  Ratios close
to 0 indicate users who maintain an up-to-date device with multiple versions for
each app.  We consider the following four update rate levels (i.e., bins) based
on the inverse of the ratio: \textit{High} \mbox{($R \le 0.25$)},
\textit{Medium} \mbox{($0.25 < R \le 0.50$)}, \textit{Low} \mbox{($0.50 < R \le
0.75$)}, \textit{Very Low} \mbox{($ R > 0.75$)}.  The results reflect a steady
risk increase as the update rate decreases.  Frequent software updates
can lower the risk of a malicious encounter by over 50\%. 

\paragraph{Distribution.} A recent study measured that 3.2\% of the apps
installed from alternative markets were PUA or malware~\cite{andropup}.  We
include as a feature the fraction of apps installed from alternative markets so
that it can be compared with other feature classes.  Indeed, this feature
represents one of the most indicative risk factors.  Downloading more than 75\%
of apps from alternative markets increases the risk of a malicious encounter 12
times. 
We also examine a second feature which counts the number of alternative markets
in the device.
We find negligible difference when comparing devices with zero alternative
markets (i.e., only \play) and those with a single alternative market.  This
indicates that the often pre-installed device vendor market does not
significantly increase user risk.  On the contrary, when more than four
alternative markets are installed, the risk increases 12 times indicating the
existence of alternative markets with much higher risk.

\paragraph{Reputation.} To capture reputation we use the app prevalence, i.e.,
the number of devices where the app has been installed.  We uniformly split the
prevalence interval $[1, max(prevalence)]$ into 4 bins and assign them
prevalence labels \textit{Very High, High, Medium and Low}.
The results show that using popular apps lowers the risk of a malicious
encounter, but the decrease is very small.  Compared to other feature classes,
app reputation correlates less with the chance of a malicious encounter.

\paragraph{Geography.} Prior work has shown that the number and types of
malicious software that hosts encounter greatly vary across
countries~\cite{ppi,yen2014epidemiological,simoiu2020targeted,botacin2021one}.
We examine two geographical features: the number of countries each device
connects from and the user's continent.
The number of countries a user connects from increases the risk in all three
models.  In particular, users connecting from more than 5 countries have a risk
of being exposed to malicious apps over two times higher than users always
connecting from the same country.  Such users may be exposed to threats that are
specific to some geographical locations such as Brazilian banking
trojans~\cite{botacin2021one} or malware that may not be distributed to specific
countries (e.g., Russia)~\cite{ppi}.  It is worth noting that such users may
include privacy-sensitive users that use a virtual private network (VPN) to
connect to the Internet, and due to the VPN usage may be exposed to threats they
may not otherwise encounter.

Finally, we examine risk differences when connecting from different continents.
The reference bins for each continent are users that do not connect from that
continent.  Africa and South America are the two continents with a higher risk
than their baseline.  Previous work has measured higher malware encounter risk
for African desktops~\cite{mezzour2015empirical}, which we confirm as well for
Android devices.  On the other hand, there is a positive trend for users
connecting from Oceania, North America, and Europe, whose risk is 40\% to 60\%
lower than their baseline.

\vspace{.2cm} \noindent \fbox{\begin{minipage}[c]{.97\columnwidth}
	\textbf{Summary}: The diversity of publishers signing the apps and the use
	of alternative markets are the highest risk factors for Android users.
	Other factors such as app update rate, device usage, app prevalence, and
geographical location are also important to define Android user risk.
Furthermore, there are significant risk differences among users in different
bins with respect to the same factor.  \end{minipage}}

\section{Profile-Based Risk Analysis} \label{sec:behavioral}

As shown in Section~\ref{sec:systematic}, there are significant differences in
risk among users with respect to the same risk factor.  In this section, we
investigate whether such risk differences may be due to user interests, i.e.,
users with common interests may share similar risk behaviors. 
To this end, we
propose a profile-based risk analysis that first groups users based on 
shared interests and then examines differences between selected user profiles.
Section~\ref{sec:clustering} details the user profile creation and
Section~\ref{sec:profiling} the profile-based risk analysis.

\begin{table}[!t]
\setlength\tabcolsep{1.5pt}
\small
\centering
\caption{App category prevalence. 
Fraction of devices with at least one app from the category;
fraction of all apps available in Play from this category;
fraction of malicious APKs in the category; and
category threshold.}
\label{tbl:categories}
\begin{tabularx}{\columnwidth}{Xrrrrr}
\toprule

		\multirow{2}{*}{\textbf{Category}}&
		\multirow{2}{*}{\textbf{Devices}}&
		\multirow{2}{*}{\textbf{Apps}}&
		\multicolumn{1}{c}{\textbf{Malicious}}&
		\multirow{2}{*}{\textbf{Threshold}}\\
		&
		&
		&
		\multicolumn{1}{c}{\textbf{APKs}}
		&\\
\midrule
Communication                   & 95.37\% & 0.50\% & 0.66\% &	 9 \\
Tools                           & 94.48\% & 1.45\% & 1.89\% &	16 \\
Productivity                    & 88.61\% & 0.84\% & 0.81\% &	12 \\
Entertainment                   & 81.72\% & 1.41\% & 1.58\% &	 8 \\
Social              			& 77.98\% & 0.46\% & 3.61\% &	 5 \\
Music and audio     			& 77.90\% & 1.36\% & 1.58\% &	 6 \\
Shopping            			& 75.49\% & 0.70\% & 0.25\% &	 8 \\
Finance             			& 73.29\% & 0.76\% & 0.38\% &	 8 \\
Travel and local    			& 67.12\% & 0.78\% & 0.40\% &	 7 \\
Video players       			& 66.59\% & 0.15\% & 1.96\% &	 4 \\
Game                			& 65.84\% & 2.97\% & 4.07\% &	14 \\
Photography         			& 65.66\% & 0.47\% & 3.12\% &	 5 \\
News and magazines  			& 65.11\% & 0.52\% & 0.63\% &	 5 \\
Lifestyle           			& 63.51\% & 1.48\% & 0.90\% &	 6 \\
Maps and navigation 			& 54.53\% & 0.34\% & 0.75\% &	 4 \\
Books and reference 			& 50.89\% & 1.18\% & 3.39\% &	 4 \\
Health and fitness  			& 49.75\% & 0.86\% & 1.37\% &	 4 \\
Business            			& 46.11\% & 1.60\% & 0.34\% &	 4 \\
Food and drink      			& 39.96\% & 0.71\% & 0.27\% &	 4 \\
Education           			& 38.37\% & 2.29\% & 2.18\% &	 4 \\
Personalization     			& 35.94\% & 1.05\% & 2.02\% &	 3 \\
Weather             			& 34.96\% & 0.09\% & 0.83\% &	 3 \\
Sports              			& 25.36\% & 0.45\% & 0.91\% &	 3 \\
Medical             			& 17.42\% & 0.38\% & 0.65\% &	 2 \\
Auto and vehicles   			& 15.71\% & 0.22\% & 0.88\% &	 2 \\
Comics              			& 12.12\% & 0.05\% & 1.57\% &	 2 \\
House and home      			&  9.73\% & 0.14\% & 0.45\% &	 2 \\
Libraries and demo  			&  8.41\% & 0.06\% & 0.81\% &	 2 \\
Art and design      			&  7.24\% & 0.18\% & 5.87\% &	 2 \\
Events              			&  3.62\% & 0.15\% & 0.10\% &	 2 \\
Dating              			&  2.65\% & 0.05\% & 0.46\% &	 2 \\
Parenting           			&  2.51\% & 0.04\% & 0.91\% &	 2 \\
Beauty                          &  2.30\% & 0.13\% & 1.38\% &	 2 \\
\bottomrule
\end{tabularx}
\end{table}

\subsection{Creating the Profiles} \label{sec:clustering}

Prior work has inferred user interests 
by applying unsupervised machine learning (ML) 
to cluster users with similar installed apps~\cite{seneviratne2015your,zhao2016mining,seneviratne2014predicting}.
Unfortunately, the black-box nature of ML techniques 
makes it difficult to understand what type of user each cluster represents.
Thus, those works typically analyze selected clusters for 
which the authors can come up with an easy-to-interpret explanation. 
In contrast, we assume that users that install multiple apps from 
the same category in \Play (e.g., \texttt{Games}, \texttt{Finance}) 
have a shared interest in the category and thus an easy-to-interpret 
common profile (e.g., gamer, investor). 
To determine if a device exhibits an interest in a category, 
we select a threshold number of installed apps corresponding to a device
being above the 10th percentile of all devices installing apps
from the category.
For example, 10.02\% of devices have installed at least 13 games
while 9\% of devices have installed at least 14 games.
Thus, 14 is selected as the threshold for the \texttt{Game} category as
a device with at least 14 games installed is in the top 10\%
of its category, and thus its owner is likely a gamer.

Table~\ref{tbl:categories} shows the prevalence of each \Pl category in our
dataset using two metrics: the percentage of devices with at least one app
installed from the category and the fraction of all apps available in \Play that
belong to the category.  The third column reports the fraction of malicious APKs
and serves as an indicator of how much a category may be
targeted by malicious developers.
The rightmost column captures the threshold for the category.
For the smallest categories 
(i.e., \texttt{Dating}, \texttt{Parenting}, \texttt{Beauty}) 
the threshold would be one app, which would be too sensitive to noise. 
For these small categories, we raise the threshold to two. 
For the interested reader, Table~\ref{tbl:profilingPercentage} in the Appendix
reports the percentage distribution of devices installing apps 
for each category, which was used for selecting the thresholds. 

We build \emph{single-category} profiles for users with a single interest
and \emph{multi-category} profiles for users with multiple interests. 
We first build 33 single-category profiles, 
one for each Play category.
Devices in a single-category profile exhibit only interest in that category
(i.e., other category thresholds are not reached).
We handle users with multiple interests 
by building \emph{multi-category} profiles. 
In particular, we build 528 two-category profiles, 
each for a combination of two categories.
For example, the \texttt{Games--Finance} profile captures devices
that are above the 10th percentile of \texttt{Games} installations and 
also of \texttt{Finance} apps installations, but not above the 10th percentile 
for any other category.
The more shared interests in a profile, the less devices satisfying the profile.
We avoid building profiles with more than two categories since they would 
represent too few devices ---e.g., there are only 8 two-category profiles with
more than 10k devices--- and would be more difficult to interpret.
Instead, we create an additional \texttt{Mixed profile} 
that consists of devices with more than two interests. 
We also build an additional \texttt{Average users} profile 
that consists of devices with an average
(i.e., within one standard deviation from the mean)
number of apps in each category and use it as a baseline in the analysis.

In total, our analysis covers 563 profiles.
We assign a single-category profile to 20.7\% of the devices 
in the filtered dataset and 
a two-category profile to 11.3\%. 
The average users profile covers 7.0\% of the devices and 
the mixed profile another 5.5\%. 
The remaining 55.5\% of devices are not profiled and thus 
are excluded from the profile-based experiments.

One caveat in our profiles is that only 53\% of all apps in the 
filtered dataset are available in \Play.
Thus, we may miss user interests that only manifest in uncategorized apps.
However, the vast majority of devices has apps from \Play with 
a mean of 47 (median of 37) Play apps per device.
Figure~\ref{sec:appendix-fig:gplay_dev_distr} in the Appendix shows the whole
distribution of \Pl apps per device.

\begin{table*}[!t]
	\small
	\centering
	\caption{Malicious-app encounter rate of average users, mixed profile,
	and top-10 profiles sorted by malicious app encounter rate among profiles with at least 10k devices.}
	\label{tbl:profileMalicious_top10}
	\begin{tabular}{lr|rrr|rr}
\toprule
		\multirow{2}{*}{\textbf{Profile}}&
		\multirow{2}{*}{\textbf{Size}}&
		\multicolumn{1}{c}{\textbf{Encounter}}&
		\multicolumn{1}{c}{\textbf{Malware}}&
		\multicolumn{1}{c}{\textbf{PUA}}&
		\multicolumn{1}{c}{\textbf{Malicious APKs}}&
		\multicolumn{1}{c}{\textbf{Malicious Encounters}}\\
		&
		&
		\multicolumn{1}{c}{\textbf{malicious apps}}&
		\multicolumn{1}{c}{\textbf{only}}&
		\multicolumn{1}{c}{\textbf{only}}&
		\multicolumn{1}{c}{\textbf{from profile apps}}&
		\multicolumn{1}{c}{\textbf{from profile apps}}\\
\midrule
Social-Video Players      &   17,184 &     6,034 (35.11\%) &        7.41\% &   81.27\% &           23.42\% &    2,248 (37.26\%) \\
Entertainment-Game        &   16,998 &     5,354 (31.50\%) &        9.30\% &   77.94\% &           46.12\% &    2,909 (54.33\%) \\
Photography-Video Players &   10,993 &     3,123 (28.41\%) &       10.79\% &   78.67\% &           24.32\% &     955 (30.58\%) \\
Video Players             &  106,723 &    27,543 (25.81\%) &        8.73\% &   79.29\% &           17.25\% &    8,245 (29.94\%) \\
Education-Game            &   14,769 &     3,633 (24.60\%) &        9.33\% &   77.92\% &           27.01\% &    1,542 (42.44\%) \\
Comics-Game               &   13,549 &     3,317 (24.48\%) &        8.38\% &   81.49\% &           40.62\% &    1,646 (49.62\%) \\
Game                      &  235,434 &    48,903 (20.77\%) &       10.53\% &   79.22\% &           34.97\% &   21,100 (43.15\%) \\
\rowcolor{red!15}Mixed                     &  470,383 &    93,407 (19.86\%) &       10.92\% &   79.70\% &                   &                   \\
Books-Education           &   12,024 &     2,267 (18.85\%) &       12.31\% &   77.06\% &            6.63\% &     297 (13.10\%) \\
Social                    &   99,176 &    18,497 (18.65\%) &       10.96\% &   81.68\% &            1.62\% &     434 (2.35\%)  \\
Business-Finance          &   16,242 &     3,000 (18.47\%) &       14.10\% &   78.60\% &           12.88\% &     410 (13.67\%) \\
\rowcolor{red!15}Average users             &  599,483 &    51,885 (8.65\%)  &       12.01\% &   73.14\% &                   &                   \\
\midrule
Overall average 		  &    6,846 &     	 990 (14.55\%) &       10.67\% &   79.20\% &           13.72\% &     151 (16.55\%) \\
\bottomrule
\end{tabular}
\end{table*}

\subsection{Profile-based risk} \label{sec:profiling}
We start our profile-based analysis by examining the 
malicious app encounter rate
in each profile. Table~\ref{tbl:profileMalicious_top10} 
lists the top-10 profiles sorted by malicious app encounter rate,
(among those with at least 10k devices) 
and the same data for the \texttt{Average users} and \texttt{Mixed} profiles. 
For each profile, the table shows the number and fraction of
devices that encounter a malicious app (malware or PUA), 
as well as the fraction that only encounter malware and 
that only encounter PUA. 
The right part of the table shows the fraction
of malicious APKs from the same categories of the profile, and the number
and fraction of devices with at least one encounter due to apps in the
profile categories, e.g., the fraction of devices from the
\texttt{Entertainment-Game} profile with encounters due to either games or entertainment
apps. The last row of the table
provides aggregated statistics for all 563 profiles including those that have
less than 10k users.
Table~\ref{tbl:profileMalicious_10K} in the Appendix contains 
similar statistics extended to all the profiles with at least 10k devices.
We also release in an anonymized repository the complete statistics
for all the 563 profiles~\cite{anonymized_malware}.

The average malicious app encounter across all profiles is 14.55\%,
almost two times higher than the one of the \texttt{Average users} (8.65\%). 
Similarly, the malicious app encounter rate for the top-10 profiles of
Table~\ref{tbl:profileMalicious_top10} is
more than double the one of the \texttt{Average users},
i.e., from 35.11\% for \texttt{Social-Video Players} to 
18.47\% for \texttt{Business-Finance}. 
On the other side of the spectrum, 
profiles such as \texttt{Comics} have much fewer encounters (5.41\%)
than the \texttt{Average users} (8.65\%),
even though the number of devices in the profile is considerably high (36.4k).
As shown in Table~\ref{tbl:categories}, this can be explained by the fact
that \texttt{Comics} apps are less often detected as malicious with respect to other
categories, such as \texttt{Game}, \texttt{Art and design}, and
\texttt{Social}. Nevertheless, the profile \texttt{Comics-Game} reveals a much
higher encounter rate (24.48\%) than the overall average (14.55\%), underlining 
the fact that users' interests in multiple categories 
can amplify their risk.
These results highlight that one-size-fits-all
cybersecurity solutions might not be suitable for everyone since users with
specific interests (e.g., games, video, social) may be exposed to
significantly higher risks.  
In terms of the threat class, 
all profiles are affected by PUA roughly seven to eight
times more than by malware.

\begin{table*}
	\small
	\centering
	\caption{Top-4 categories responsible for malicious-app encounters for
	average users, mixed profile, and top-10 profiles sorted by malicious app encounter rate among profiles with at least 10k devices.}
	\label{tbl:threatSources_top10}
\begin{tabular}{lrrrr}
\toprule

		\multirow{2}{*}{\textbf{Profile}}&
		\multicolumn{4}{c}{\textbf{Top-categories}}\\
		&
		\multicolumn{1}{c}{\textbf{1st}}&
		\multicolumn{1}{c}{\textbf{2nd}}&
		\multicolumn{1}{c}{\textbf{3rd}}&
		\multicolumn{1}{c}{\textbf{4th}}\\
\midrule
Social-Video Players      &  Video Players (50.46\%) &           Tools (21.8\%) &            Game (5.68\%) &  Entertainment (3.49\%) \\
Entertainment-Game        &           Game (55.54\%) &  Entertainment (17.99\%) &           Tools (7.98\%) &          Music (5.18\%) \\
Photography-Video Players &  Video Players (26.33\%) &    Photography (24.07\%) &          Tools (17.84\%) &          Music (6.74\%) \\
Video Players             &   Video Players (46.9\%) &          Tools (19.27\%) &            Game (7.53\%) &          Music (4.44\%) \\
Education-Game            &           Game (56.05\%) &          Tools (10.62\%) &   Entertainment (6.03\%) &  Video Players (5.08\%) \\
Comics-Game               &           Game (67.02\%) &   Video Players (8.61\%) &           Tools (7.51\%) &          Music (5.12\%) \\
Game                      &           Game (64.04\%) &           Tools (8.02\%) &   Entertainment (7.56\%) &  Video Players (5.42\%) \\
\rowcolor{red!15}Mixed                     &          Tools (25.37\%) &           Game (13.51\%) &  Video Players (10.00\%) &          Music (8.35\%) \\
Books-Education           &          Tools (23.64\%) &            Game (11.2\%) &      Education (10.47\%) &  Video Players (9.38\%) \\
Social                    &           Tools (31.0\%) &  Video Players (22.05\%) &            Game (9.83\%) &          Music (5.06\%) \\
Business-Finance          &          Tools (31.08\%) &         Finance (13.2\%) &       Business (11.58\%) &  Video Players (8.72\%) \\
\rowcolor{red!15}Average users             &          Tools (30.35\%) &           Game (16.83\%) &  Video Players (16.23\%) &  Entertainment (5.90\%) \\
\midrule
Overall average           &          Tools (22.19\%) &           Game (12.58\%) &           Music (7.27\%) &  Video Players (7.18\%) \\
\bottomrule
\end{tabular}
\end{table*}

Some app categories may be targeted more frequently by malicious developers
than others, and thus have a larger fraction of malicious apps.  For example,
the last column of Table~\ref{tbl:categories} shows that 5.87\% and 4.07\% of
all the APKs in the \texttt{Art and design} and \texttt{Game} categories are
flagged as malicious.  Prior work also
identifies the \texttt{Game} category as the one in \Play 
that hosts the most malicious
APKs~\cite{wang2019rmvdroid}.  Thus, users whose interests fall into 
those highly abused categories, have a higher risk of encountering malware.
Indeed, according to the last column in
Table~\ref{tbl:profileMalicious_top10}, 
more than 50\% of users in the \texttt{Entertainment-Game} profile encounter
malware from one of those two categories. 
Overall, 5 of the top-10 profiles in Table~\ref{tbl:profileMalicious_top10},
have encounters from malicious apps in their categories at least twice 
higher than the average of all profiles (16.55\%), 
indicating that users in those profiles have a higher
intrinsic risk due to their interests.

We further investigate the source of the risk by identifying 
for each profile the top-4 application
categories where the most number of malicious apps come from.
In Table~\ref{tbl:threatSources_top10}, we report the results for the same set of 12 profiles identified in
Table~\ref{tbl:profileMalicious_top10}. We instead tabulate results for all the profiles with at least 10k users in
Table~\ref{tbl:threatSources_10K} of the Appendix, and we release the full
measurement on all the profiles in an anonymized
repository~\cite{anonymized_malware}.
On average, the biggest offenders are
\texttt{Tools}, \texttt{Game}, \texttt{Music}, and \texttt{Video Players}.  The
\texttt{Tools} category is the most prevalent threat source for 5 of of the
profiles in Table~\ref{tbl:threatSources_top10}, and among the top-4 for all profiles with more than 10k
devices (Table~\ref{tbl:threatSources_10K}).  Other categories
frequently appearing in the top-4 are \texttt{Game} (86\% of the profiles)
and \texttt{Video Players} (59\%).
However, for some profiles, their defining categories contribute the most.
In particular, 211 (37\%) over all generated profiles 
and 12 of the 35 profiles with at least 10k users, 
have their own category as the top threat source. For example, the
\texttt{Health and Fitness}, \texttt{Photography}, \texttt{Weather}, \texttt{Finance}, and
\texttt{Shopping} profiles have a higher risk of malware encounters due to
their interests, rather than due to the categories of apps that are more likely
to be bringing malicious apps compared to the others.

\begin{table*}[!t]
    \centering
    \small
    \caption{Comparison of the odds ratios of installing PUA and/or malware
	between the baseline and the last bin for the average users, mixed profile,
	and the top-10 user profiles among those with at least 10k users sorted by
	malware encounter rate.}
	\label{tbl:profileodds_top10}
	\setlength{\tabcolsep}{1pt}

    \begin{tabular}{lrrrrrrrrrrrr}
    \toprule
		\multirow{2}{*}{\textbf{Feature}}&
		\multicolumn{1}{c}{\textbf{Social}}&
		\multicolumn{1}{c}{\textbf{Entertainment}}&
		\multicolumn{1}{c}{\textbf{Photography}}&
		\multicolumn{1}{c}{\textbf{Video}}&
		\multicolumn{1}{c}{\textbf{Education}}&
		\multicolumn{1}{c}{\textbf{Comics}}&
		\multirow{2}{*}{\textbf{Game}}&
		\multirow{2}{*}{\textbf{Mixed}}&
		\multicolumn{1}{c}{\textbf{Books}}&
		\multirow{2}{*}{\textbf{Social}}&
		\multicolumn{1}{c}{\textbf{Business}}&
		\multicolumn{1}{c}{\textbf{Average}}\\

		&
		\multicolumn{1}{c}{\textbf{Video Players}}&
		\multicolumn{1}{c}{\textbf{Game}}&
		\multicolumn{1}{c}{\textbf{Video Players}}&
		\multicolumn{1}{c}{\textbf{Players}}&
		\multicolumn{1}{c}{\textbf{Game}}&
		\multicolumn{1}{c}{\textbf{Game}}&
		&
		&
		\multicolumn{1}{c}{\textbf{Education}}&
		&
		\multicolumn{1}{c}{\textbf{Finance}}&
		\multicolumn{1}{c}{\textbf{users}}\\

\midrule
		Apps  			&  1.55  				&  1.02  &  \underline{0.71}  	&  1.19  &  1.11  			&  1.06  &  1.48  &  \underline{0.69}  	&  1.68  			&  1.27  			&  \underline{0.86}  	&  1.39\\
		Activitydays  	&  1.88  				&  1.33  &  \underline{0.98}  	&  1.66  &  1.03  			&  1.31  &  1.25  &  1.32  				&  1.04  			&  1.46  			&  1.13  				&  1.50\\
		Signers  		&  1.42  				&  3.91  &  3.60  				&  2.53  &  4.15  			&  3.47  &  3.05  &  3.26  				&  2.53  			&  2.38  			&  3.28  				&  3.64\\
		AppCateg.  		&  0.79  				&  0.65  &  0.41  				&  0.42  &  0.78  			&  0.56  &  0.70  &  0.47  				&  0.67  			&  0.52  			&  0.85  				&  0.41\\
		Updaterate  	&  1.15  				&  1.87  &  1.48  				&  2.21  &  1.79  			&  3.23  &  2.33  &  1.70  				&  1.43  			&  1.48  			&  1.05  				&  3.28\\
		\%AppsAM  		&  1.71  				&  2.82  &  3.72  				&  2.10  &  \textbf{33.08}  &  3.41  &  6.53  &  \textbf{26.78}  	&  \textbf{19.48}  	&  1.88  			&  11.38 				&  4.14\\
		Alt.Markets  	&  2.94  				&  2.62  &  4.41  				&  2.44  &  3.46  			&  3.63  &  2.63  &  2.36  				&  1.65  			&  1.65  			&  1.78  				&  1.28\\
		AppPrev.  		&  0.63  				&  0.77  &  0.85  				&  0.87  &  0.77  			&  0.80  &  0.75  &  0.70  				&  0.69  			&  0.48  			&  0.73  				&  0.61\\
		Countries  		&  \underline{0.40}  	&  1.17  &  1.15  				&  1.55  &  1.08  			&  1.03  &  1.09  &  1.18  				&  1.15  			&  \underline{0.30} &  1.41  				&  1.26\\
\bottomrule
\end{tabular}
\end{table*}

In addition, we perform profile-based odds ratio analysis on the same set of
selected profiles of Tables~\ref{tbl:profileMalicious_top10}
and~\ref{tbl:threatSources_top10} and report the results in
Table~\ref{tbl:profileodds_top10}. The same results for the profiles larger than
10k devices are reported in Table~\ref{sec:behavioral2:profileodds} of the
Appendix.
Similarly to
Section~\ref{sec:systematic}, we bucketize each feature into quartiles, however
here our odds ratio analysis is performed separately for each profile.  Due to
space constraints, we do not report detailed risk analysis results separately
for PUA and malware, PUA-only, and malware-only encounters.  Instead,
Table~\ref{tbl:profileodds_top10} summarizes the results comparing the
risk estimation of the second bin with the last bin.  Since the sizes of the
bins are different among profiles, it would not be accurate to do comparisons
among the absolute numbers presented in the Table.  Our goal here is to identify
trends for each profile.  Cells with underlined values correspond to opposite
trends compared to our findings in Section~\ref{sec:systematic}.  Cells with
bold values, on the other hand, represent features with significant impact on
users risk.

One of our findings in Section~\ref{sec:systematic} is that as volume-based
features (e.g., total number of apps, active days) increase, so do the
odds of a malicious encounter.  While this remains the same for most user
profiles, there are three profiles (\texttt{Photography-Video Players},
\texttt{Mixed}, \texttt{Business-Finance}) for which we observe the opposite
trend with a slight difference. We observe a
similar phenomenon for the two profiles related to social media
(\texttt{Social} and \texttt{Social-Video Players}), in which the odds of
malicious encounter decrease as the number of countries the device connected
from increases. 

This analysis further illustrated that risk profiles significantly vary for
users with particular interests.  In some cases, we observe reverse correlations
for the same features among profiles as well as dramatically different impact of
the same features on different profiles.  The most evident examples of the
latter
are the \texttt{Education-Game}, \texttt{Mixed}, and \texttt{Books-Education},
 profiles for which the
fraction of the apps installed from alternative markets can increase the risk up
to 33 times with respect to those that install less in their class.

\vspace{.2cm} \noindent \fbox{\begin{minipage}[c]{\columnwidth}
	\textbf{Summary}:

The per-profile risk analysis results suggest that the risk profiles of the 
	 investigated categories are significantly different compared to the global
	risk analysis results. Although in general they share similar trends, we
	observe opposite trends for some of the features. Furthermore, the
contribution of some of the indicators such as the number of apps installed from
alternative markets can be much more significant for some profiles while not very important for the others.  \end{minipage}}

\begin{table*}
	\centering
	\caption{Classification accuracy of whole-population and per-profile models
	tested on average users, mixed profile, and on the top-10 profiles among those
	with at least 10k users sorted by malware encounter rate.}
        \label{tbl:classification}
\small
\begin{tabular}{l|rrr|rrr|r}
\toprule
		\multirow{3}{*}{\textbf{Profile}}&
		\multicolumn{3}{c|}{\textbf{Whole-population model}}&
		\multicolumn{3}{c|}{\textbf{Per-profile models}}&
		\multirow{3}{*}{\textbf{Avg improvement}}\\
		&
		\multicolumn{1}{c}{\textbf{PUA or}}&
		\multicolumn{1}{c}{\textbf{Malware}}&
		\multicolumn{1}{c|}{\textbf{PUA}}&
		\multicolumn{1}{c}{\textbf{PUA}}&
		\multicolumn{1}{c}{\textbf{Malware}}&
		\multicolumn{1}{c|}{\textbf{PUA}}&
		\\
		&
		\multicolumn{1}{c}{\textbf{Malware}}&
		\multicolumn{1}{c}{\textbf{only}}&
		\multicolumn{1}{c|}{\textbf{only}}&
		\multicolumn{1}{c}{\textbf{Malware}}&
		\multicolumn{1}{c}{\textbf{only}}&
		\multicolumn{1}{c|}{\textbf{only}}&
		\\
\midrule
Social-Video Players      &                   50.40\% &                       64.64\% &                   48.30\% &                      82.43\% &                          78.35\% &                      69.23\% &            22.22\% \\
Entertainment-Game        &                   50.10\% &                       62.89\% &                   49.80\% &                      68.00\% &                          63.16\% &                      61.29\% &             9.89\% \\
Photography-Video Players &                   53.00\% &                       63.23\% &                   61.27\% &                      66.27\% &                          80.00\% &                      62.90\% &            10.56\% \\
Video Players             &                   48.50\% &                       50.10\% &                   53.10\% &                      73.20\% &                          82.16\% &                      67.24\% &            23.63\% \\
Education-Game            &                   45.80\% &                       70.00\% &                   51.10\% &                      74.04\% &                          62.90\% &                      69.47\% &            13.17\% \\
Comics-Game               &                   48.00\% &                       70.11\% &                   49.70\% &                      75.76\% &                          77.27\% &                      81.25\% &            22.16\% \\
Game                      &                   53.50\% &                       49.60\% &                   48.80\% &                      64.38\% &                          64.22\% &                      72.22\% &            16.31\% \\
\rowcolor{red!15}Mixed 	  &                   50.30\% &                       51.40\% &                   52.90\% &                      72.16\% &                          68.42\% &                      67.03\% &            17.67\% \\
Books-Education           &                   61.78\% &                       63.98\% &                   61.61\% &                      80.95\% &                          70.00\% &                      81.58\% &            15.05\% \\
Social                    &                   52.30\% &                       49.60\% &                   47.30\% &                      54.76\% &                          75.29\% &                      70.21\% &            17.02\% \\
Business-Finance          &                   49.60\% &                       58.67\% &                   62.33\% &                      62.12\% &                          53.85\% &                      70.59\% &             5.32\% \\
\rowcolor{red!15}Average users             &                   51.90\% &                       49.50\% &                   51.50\% &                      84.35\% &                          83.56\% &                      77.91\% &            30.97\% \\
\midrule
Overall average           &                   52.19\% &                       62.34\% &                   52.89\% &                      69.84\% &                          70.14\% &                      71.65\% &            14.74\% \\
\bottomrule
\end{tabular}
\end{table*}

\section{Multi-Feature Risk Analysis} \label{sec:prediction}

In previous sections, we have shown that particular users' interests result in
different risk profiles and that risk can significantly vary among users. We
believe our findings can assist the security community and the designers of
security solutions on providing tailored protections to users who need it the
most. In this respect, we provide a concrete example of how our profiling
strategy and the differences among profiles impact the classification task of
devices that encounter malicious apps.
To this end, we compare two machine learning (ML) classification approaches:
using a one-size-fits-all classifier trained on the whole population 
(i.e., without any profiling) and using per-profile classifiers, 
each trained on a profile's population. 
For both approaches, we create three classifiers for
identifying devices that install 
(1) any malicious app (i.e., malware or PUA), 
(2) malware, and 
(3) PUA. 

For the whole-population classifier, we use 86 features: 
15 features correspond to those discussed in Section~\ref{sec:systematic}
with the Countries feature being one-hot encoded, 
37 boolean features that capture
whether the device belongs to one of the profiles with more than 10k users
discussed in Section~\ref{sec:behavioral} and reported in
Table~\ref{tbl:profileMalicious_10K},
33 features that measure the number of apps
installed for each \Pl category in Table~\ref{tbl:categories}, and one
feature capturing the absolute number of uncategorized apps (i.e.
\texttt{Unknown}). This classifier outputs a clean state or the presence of
malicious installations for each device independently of their profile: we then
link back the device to one of the 37 profiles to analyze per-profile performance.
The per-profile classifiers use 49 features, the same as the whole-population
classifier minus the 37 boolean features that indicate the user profiles.

We select a Random Forest (RF) classifier for the experiments because this
learning method produces easier-to-interpret models and  
because prior work that builds prediction models for cyber incidents has shown
it tends to outperform other classifiers
(e.g.,~\cite{riskteller,liu2015cloudy,sharif2018predicting}). 
We tune the RF hyperparameters 
by looking at the Out of Bag (OOB) error during the training phase.  Our dataset
converges by employing $225$ tree estimators with $20$ as maximum tree depth,
and the maximum number of features to consider when splitting each branch
being the square root of the number of features. 

Since the number of devices for each profile differs, the dataset used to train
and test the whole-population model is obtained by sub-sampling each of the
profiles and by further balancing the ratio between clean and infected devices.
For each model, we sub-sample multiple times and average accuracy results.  We
use the same users selected at this step for building the per-profile models.
Table~\ref{tbl:datasetSize} in the Appendix reports the exact
dataset sizes of all models.

Table~\ref{tbl:classification} presents the classification accuracy 
for both approaches when tested against profiles of Tables~\ref{tbl:profileMalicious_top10}
and~\ref{tbl:threatSources_top10}.
Table~\ref{tbl:classification_appendix} in the Appendix is similar 
but tests all profiles with at least 10k devices. We observe a clear
improvement on identifying risky devices when the models are trained
per-profile rather than on the whole population: the average classification
accuracy for any malicious app encounter (PUA or malware) increases on average from
$52.19\%$ to $69.84\%$. This holds for all the profiles and all the
threat types, although for some profiles the impact of the profiling is much
higher than others. 
For example, the classification of \texttt{Average users} devices 
would benefit from a leap in accuracy of over 30\% compared to using
a one-size-fits-all classifier.
In a similar way, a dramatic accuracy improvement happens on the \texttt{Video Players} category where
$23.63\%$ more devices are categorized correctly. The least amount of
accuracy improvement occurs for \texttt{Productivity} (0.53\%) and
\texttt{Lifestyle} (0.38\%).
For these cases, there is alignment between the accuracy score of the 
model specifically trained for that population and the one trained on the whole
population, although the classification score is always higher in
the former case. For these profiles, the classification of clean and
devices that encounter malicious apps might be more complicated. This
strongly suggests that such populations would benefit even more from
 specific models (e.g., dividing it into sub-categories, combining them with other categories etc.). We indeed notice an
accuracy score barely above the output of a random classifier (i.e., 50\%) for
19 of the 37 profiles of Table~\ref{tbl:classification_appendix}; even worse, 
the outcome is lower than 50\% for 12 profiles when predicting the devices that
encounter malware or PUA. Once again, such findings further indicate the need of
dedicated methods for those users.

The results also show different accuracy
across profiles.  For example, \texttt{Social-Video Players} (82.43\%),
and \texttt{Books-Education} (80.95\%) achieve much higher
classification accuracies compared to \texttt{Social} (54.76\%) and
\texttt{Business-Finance} (62.12\%) profiles.
This might suggest that for some of the
identified populations the users installing malicious apps have a more
diversified set of features with respect to clean ones, thus helping the
classification task. For other profiles with lower accuracy, this difference might
not be fully captured by the features at our disposal, thus making it harder to
classify the set of clean devices from the rest.

\begin{table*}[!t]
	\centering
	\setlength{\tabcolsep}{2.7pt}
\footnotesize
	\caption{Top-5 features of the classifiers trained on the whole-population, 
	average users, mixed profile, and on the top-10 profiles sorted by
	malware encounter rate among the profiles with at least 10k users.}
	\label{sec:prediction-tbl:features}
	\begin{tabular}{crrrrrrr}
\toprule
		\textbf{\Centerstack{Feature\\rank}}&
		\textbf{\Centerstack{Unique\\model}}&
		\textbf{\Centerstack{Average users}}&
		\textbf{\Centerstack{Mixed}}&
		\textbf{\Centerstack{Game}}&
		\textbf{\Centerstack{Video\\Players}}&
		\textbf{\Centerstack{Social}}&
		\textbf{\Centerstack{Social\\Video Players}}\\
\midrule
1st &      Signers - 9.76 &     Unknown - 19.93 &    Avg prev. - 7.56 &      Unknown - 16.5 &     Unknown - 10.83 &     Unknown - 10.83 &      Unknown - 11.69 \\
2nd &      Unknown - 8.93 &    Avg prev. - 9.92 &       Signers - 6.9 &          Apps - 8.6 &      Signers - 9.02 &    Avg prev. - 8.65 &    Avg prev. - 11.16 \\
3rd &   \% apps AM - 5.61 &      Signers - 7.11 &  Update rate - 6.19 &       Signers - 6.7 &    Avg prev. - 7.32 &         Days - 6.48 &          Apps - 6.76 \\
4th &         Apps - 5.39 &  Update rate - 6.91 &         Days - 5.89 &    Avg prev. - 6.33 &         Apps - 6.97 &         Apps - 5.64 &       Signers - 6.22 \\
5th &  Update rate - 5.15 &         Apps - 5.66 &         Apps - 4.67 &  Update rate - 5.77 &  Update rate - 5.78 &      Signers - 5.19 &   Update rate - 4.64 \\
\bottomrule
\end{tabular}
\setlength{\tabcolsep}{2.6pt}
\begin{tabular}{crrrrrrr}
\toprule
		\textbf{\Centerstack{Feature\\rank}}&
		\textbf{\Centerstack{Entertainment\\Game}}&
		\textbf{\Centerstack{Business\\Finance}}&
		\textbf{\Centerstack{Education\\Game}}&
		\textbf{\Centerstack{Comics\\Game}}&
		\textbf{\Centerstack{Books\\Education}}&
		\textbf{\Centerstack{Photography\\Video Players}}&
		\textbf{\Centerstack{Overall\\average}}\\
\midrule
1st &     Unknown - 8.82 &  Update rate - 7.73 &      Unknown - 8.87 &     Unknown - 12.78 &   Avg prev. - 10.27 &          Avg prev. - 8.52 &      Unknown - 8.18 \\
2nd &     Signers - 8.57 &         Days - 6.97 &      Signers - 7.43 &      Signers - 7.07 &      Unknown - 9.85 &            Signers - 7.48 &    Avg prev. - 7.51 \\
3rd &         Apps - 7.3 &    Avg prev. - 6.78 &  Update rate - 7.21 &         Apps - 6.45 &         Days - 7.54 &            Unknown - 6.99 &       Signers - 7.1 \\
4th &        Days - 7.07 &      Unknown - 5.23 &         Apps - 6.68 &     Avg prev. - 5.9 &       Signers - 7.0 &               Apps - 5.52 &  Update rate - 5.86 \\
5th &        Game - 6.22 &         Apps - 4.82 &         Days - 6.09 &         Game - 5.85 &   \% apps AM - 6.97 &        Update rate - 5.48 &         Apps - 5.39 \\
\bottomrule
\end{tabular}
\end{table*}

\paragraph{Feature importance.} We also
analyze the contribution of individual features for all the trained classifiers.
We rank feature importance by assessing the Mean Decrease in Impurity (MDI) and
tabulate the average of their relative importance in the three cases (e.g.,
considering malware, PUA, and malware or PUA) in
Table~\ref{sec:prediction-tbl:features} for the top profiles and
Table~\ref{sec:prediction-tbl:features:appendix} in the Appendix
for all the profiles with more than 10k devices.
For the whole population classifier, the most contributing
features are: signers, uncategorized apps, the percentage of apps from
alternative markets, the number of apps, and the update rate. For the per-profile
classifiers, important features change for each profile. On average, the number
of unclassified apps (i.e., \texttt{Unknown}) is the most important feature,
followed by the one reflecting the average application prevalence (i.e.,
\texttt{Avg prev.}), the diversity
of signers (i.e., \texttt{Signers}), the frequency of updates (i.e.,
\texttt{Update rate}), and the number of installed apps (i.e.,
\texttt{Apps}). The relevance of different features that are not so
characteristic for other profiles explains why the classification
accuracy significantly improves when employing a specific model
rather than a unique solution that has to generalize on multiple profiles.
While the order of the list and magnitude of each feature 
importance differs significantly
among different profiles, we observe that in general the volume- and
diversity-based features (e.g., signers, app category and number, days of activity, and average
prevalence) remain on the top of the list across profiles.  

\vspace{.2cm} \noindent \fbox{\begin{minipage}[c]{\columnwidth}
	\textbf{Summary}: The multi-feature risk analysis unveils important
	differences among profiles when attempting to distinguish clean devices from
	those that install malicious apps. For some of the profiles, the
	classification accuracy shows a substantial improvement when employing a
	dedicated model, often due to the difference importance of features compared
	to the other profiles.  Overall, volume- and diversity-based features are
the most relevant indicators across profiles.  \end{minipage}}

\section{Discussion and Limitations} \label{sec:discussion} 

\paragraph{Ethical considerations.} The data in the app installation logs comes
from human subjects.  It is collected by an AV engine installed on Android
devices belonging to real users.  The telemetry only includes users who
voluntarily install the product, accept the company’s privacy
policy~\cite{nortonPolicy}, and opt-in to share their data.  Users can revoke
their consent at any time by modifying the app settings.  Data is anonymized on
the device before being sent to a central system.  The telemetry data only
identifies users through numeric anonymized identifiers that do not enable us to
trace back to the originating device or its user.  We do not attempt to
deanonymize users or to profile specific users focusing instead on aggregated
risk profiles.

\paragraph{How our results help the security community.}
The mid-term goal of our work is to promote more personalized security. 
But, our results can also be applied in the short term. 
One concrete application is to adjust the AV engine settings 
according to the risk profile. 
For example, the frequency of the scan and the scan level 
(i.e., whether all files are scanned or only a subset)
could be increased for devices exhibiting profiles with a high risk 
of malicious encounters.
Another application is to deactivate unneeded security warnings,
which create habituation that desensitizes users to other 
important warnings~\cite{krol2012don}.
For example, if a device has a \texttt{Parenting} profile, the AV engine
could avoid notifying the user about parental-control apps installed, 
while keeping the notifications for other users since such apps 
can be abused as stalkerware~\cite{stalkerware}.
The profiles could also be used to provide personalized security ``nudges'',
i.e., warnings that try to entice a user action without forcing the user.
In Android, AV engines cannot remove detected malicious apps by themselves 
(unless pre-installed with the required privilege) and 
thus nudge users to remove the detected apps.
However, such nudges are not currently effective, 
as recent work shows that Android malware and PUA
is only removed on average 24 days after the 
first detection~\cite{shen2022large}. 
Previous work has shown that personalization can improve 
security nudges~\cite{harbach2014using,liu2016follow,wisniewski2017making} 
and the inferred profiles could be used towards that goal.

\paragraph{Device-user mapping.}
A user could own multiple devices
and a device could be shared among multiple users. 
When a user has multiple devices in the dataset, 
our analysis will assign profiles separately to each device, 
and if the devices are used for different tasks, 
the analysis will infer different profiles for each device, 
e.g., one for the gaming device, another for the work device. 
Since the device belongs to a single user, 
the inferred profile will be correct, but we will miss the fact that the user 
has other interests that manifest in the other device.
Prior work has shown that mobile phone sharing is common 
within households~\cite{bodker2012poetry,matthews2016she} and 
in certain geographical locations such as 
South Asia~\cite{ahmed2013ecologies,sambasivan2018privacy,ahmed2019everyone,al2021we} 
and Africa~\cite{murphy2011my}.
When a device is shared among multiple users, 
and each user installs its own apps, 
the inferred profiles may not uniquely correspond to one user,
e.g., an inferred \texttt{Games--Finance} profile may be due to 
one user of the device being a gamer and another being an investor.
It is also possible that a profile is inferred due to the 
accumulation of partial interests from multiple users, 
especially for app categories with low thresholds.
We acknowledge this limitation, although it is worth noting that 
only 0.25\% of devices in the filtered dataset come from 
Bangladesh, Pakistan, and Kenya and that the security vendor clients are 
biased towards North America, Europe, and Japan,
where device sharing is not as popular as in South Asia and Africa.
In addition, devices shared in households may have a dominant owner that 
the inferred profiles correspond to.

\paragraph{Profile interpretability vs coverage.}
In contrast to previous ML clustering approaches our profiles 
are easy to interpret
since they capture devices exhibiting one or two interests based on 
the categories of installed apps.
The downside is that we can only profile one third of the devices.
We favor interpretability over device coverage because we believe  
that personalized security requires a good understanding of the user, 
but personalization does not necessarily need to be applied 
to every user (e.g., average users).

\paragraph{Selection bias.} 
Our dataset comes from a single \vendor and thus has a
selection bias towards the geographical distribution of the \vendor's users.
While the dataset has great geographical visibility (i.e., users from 243
countries) some countries like China or Indonesia may be under-represented with
respect to their population.
The dataset only covers users that have an AV engine installed and have
explicitly enrolled in data sharing.  Other users, e.g., those that do not
invest in security solutions or those that decline the collection due to privacy
concerns, may have different risk profiles that we cannot analyze.
Our data captures only a partial view of user behavior.  It captures whether the
user installed an app, but does not capture how frequently the app is used, and
it does not cover other user behaviors such as Web browsing.  

\paragraph{Analysis biases.}
Our VT querying, app categorization, and app filtering 
may have introduced other biases.
First, due to VT API limits, we could not query all apps in the dataset.
This prevents us from detecting all malicious encounters 
and thus we underestimate the malicious encounter rate for each profile.
However, due to the size of our dataset and the query process, 
we believe that the relative encounter rate among profiles 
should not be significantly affected.
Second, we categorize over 500K apps available on \Play on February 2020.
But, these are only 53\% of all apps in the filtered dataset.
This prevents us from detecting additional profiles that may only manifest 
through apps distributed through alternative vectors.
Still, devices in the filtered dataset have a mean of 47 categorized apps
(median of 37), allowing us to identify a profile for one third of the 
8.9M devices.
We plan to address how to categorize apps not in \Play in future work.
Finally, we filter apps that are not installed from markets 
as a proxy for removing apps whose installation is not rooted on a user
decision.  Such filtering may incorrectly remove user-installed apps when a
market is missing in our list of 122 market package names and if the user
installed the app through other vectors such as web downloads or the ADB bridge.
However, after the filtering, our dataset is still large
containing 8.9M devices and 2.2M apps.

\section{Related Work}
\label{sec:related}

\paragraph{Security risk analysis and prediction}.
Multiple works have focused on the prediction of security incidents on different platforms.
Canali et al.~\cite{canali2014effectiveness} analyzed the browsing behavior of 160k 
users and crafted 74 features for predicting users that visit malicious websites.
Sharif et al.~\cite{sharif2018predicting} proposed a similar system,
that instead of long-term prediction,
is capable of predicting exposure to malicious content seconds before the actual
exposure. Their system is evaluated using HTTP traffic from 20.6K users.

Shen et al.~\cite{shen2021andruspex} proposed a system for predicting the installation
of malicious applications on Android by observing the apps installed by the user
as well as from users with similar behaviors.
Their system is evaluated on a dataset of 1.8M real Android users.
%
Other works focus on the prediction of malware encounters on Windows.
Bilge et al.~\cite{riskteller} propose RiskTeller, a system that predicts
malware encounters on Windows enterprise hosts using 89 features.
The system is evaluated on telemetry data collected from 600K machines from 18 enterprises.
This study also uses classifiers for predicting malicious encounters on Android using similar features.
But, our goal is not to improve the state of the art in terms of prediction accuracy.
Instead, we build various classifiers for users of specific profiles (e.g., gamers,
investors) and do comparisons among them as well as against a generic classifier
to identify the differences in the importance of features. 

Other research lines analyze the various risk factors affecting users'
security posture.
Simoiu et al~\cite{simoiu2020targeted} analyzed 1.2 billion phishing and 
malware attacks against Gmail users and explore risk correlations 
for six factors, country of access,
user age, security posture, prior risk exposure, type of device, and 
email activity.
Similar to this study, we investigate correlations of 10 risk factors on 
8.6M Android users.
Kotzias et al.~\cite{andropup} analyzed the unwanted apps prevalence 
(malware and PUA) and distribution on 12M Android devices.
Their study showed that Play market is the largest app distribution vector of both benign and unwanted apps, and installations from alternative markets are on average five times riskier than Play market with varying risks depending on the alternative market.
Our work confirms that apps installed from alternative sources 
rather than the official Play market represent one of the most indicative 
risk factors of our model.
However, in this work, in addition to distribution-related risk factors, we
analyze five more types of risk factors related to volume, app diversity,
updates, reputation, and geography.

\paragraph{Security posture of at-risk populations.}
Arguments in favor of specialized solutions for vulnerable and at-risk
populations have shifted the security research community interest towards that direction~\cite{onesizefitsall}.
Warford et al.~\cite{warford2022sok} recently analyzed prior work on the 
digital-safety experiences of at-risk populations
and developed a framework of contextual risk factors and protective practices.
Consolvo et al.~\cite{thomas21usenix} conducted a qualitative analysis of the
security issues of 28 people involved with the US political campaigns and provided recommendations on security improvements.
Similarly, McGregor et al.~\cite{190976} analyzed the security practices of 15
journalists from the US and France via semi-structured interviews.
Other works have followed a similar qualitative approach to  analyze the security and privacy needs of vulnerable populations like human trafficking survivors~\cite{235451}, victims of intimate partner violence~\cite{10.1145/3359304}, residents at homeless shelters~\cite{48171}, undocumented immigrants~\cite{10.1145/3173574.3173688}, and proposed recommendations for improving the ways that such groups interact
with technology.
At last, some studies focused on the needs of specific age groups.
Frik et al.~\cite{238333} identified common threat models, misconceptions, and mitigation strategies for older adults (65+).
Wisniewski et al.~\cite{10.1145/2858036.2858317} performed an analysis of the online risk experiences of 68 teens for two months.
Lastdrager et al.~\cite{10.5555/3235924.3235943} focused on  
training children of Dutch primary school to distinguish phishing from non-phishing pages.
Inspired by these studies, we perform an empirical quantitative analysis of risk
factors for different profiles of users and provide further evidence that
one-size-fits-all cybersecurity solutions might not be suitable for 
everyone since users with specific interests (e.g., games, social, education) 
may be exposed to significantly higher risks. 

\paragraph{Android User profiling.}
An extensive body of prior work attempts to profile Android users for different purposes~\cite{zhao2019user}.
These studies analyze the smartphone interactions of users to learn various characteristics about them like demographic attributes such as gender and age~\cite{ying2012demographic, zhao2019gender, seneviratne2014predicting, seneviratne2015your}, personal interests~\cite{zhao2016mining}, psychological status like the users stress levels~\cite{gao2014predicting}, and lifestyle related information~\cite{jesdabodi2015understanding}.
It is common for those studies to do their profiling using
the list of installed apps~\cite{seneviratne2015your,zhao2016mining,seneviratne2014predicting} but also using app usage information like the number of times an app is launched or the amount of time spent on each app~\cite{do2011smartphone}.
In our study, we also use the list of installed apps for creating user profiles.
But, instead of predicting specific characteristics of users,
we perform a comparison of the security posture among profiles and the most
important risk factors.

\section{Conclusions} \label{sec:conclusions} 
Over the past years, a growing number of researchers have studied 
the digital-safety needs of particular user communities, 
highlighting the inadequacy of current one-size-fits-all security solutions.
These works are usually qualitative and limited in size.
To fill this gap, we perform a quantitative risk analysis that compares
the risk of encountering malicious apps (malware and PUA) 
for different profiles of Android users.
Leveraging telemetry data of a popular \vendor, 
we analyze the impact of different risk indicators.
We build easy-to-interpret user profiles and 
show how there exist risk differences across users due to their interests.
Finally, we identify important differences among profiles when 
attempting to distinguish clean devices from those that install malicious apps. 
We hope our findings can motivate the security community
to address the security and privacy needs of particular at-risk communities 
and offer more personalized solutions.

\bibliographystyle{plain}
\bibliography{biblio}
\appendix
\section{Additional Results}
\label{sec:additionalResults}

\subsection{Best vs Worst Devices}\label{appendix:best_worst}
In this section, we validate the importance
of the risk indicators identified in section~\ref{sec:systematic} by performing a comparison between
the best and the worse-in-class users in terms of malicious apps encounters. In the
following paragraphs, we describe the selection criteria for each group, the
insights unveiled when composing the two classes, and the results of the analysis.

Identifying devices with the best devices in our dataset requires more criteria beyond
simply selecting devices with no malicious encounters.
We decide to use the first 8 risk factors as criteria and select users from the
riskiest buckets for each factor in the GLMs
(Table~\ref{sec:systematic-tbl:glm}). 
The rationale is that users who have risky behaviour but do not encounter any
malicious apps may represent users that are very security conscious.
We observe that there is no single device that falls in the riskiest bucket of
all risk factors.  To investigate which are the most critical features that
filter out the majority of devices, we compute and report in
Table~\ref{sec:systematic-tbl:cleanvsdirty} the percentage of clean and infected
devices after applying the riskiest condition for each feature.  For example, we
compute the percentage of devices with more than 102 installed apps, since this
value resulted to have the highest odds ratios
(Table~\ref{sec:systematic-tbl:glm}).  We find out that the number of
alternative markets and the percentage of applications downloaded from
unofficial markets are very discriminative for the two sets, as only 214 and 19K
devices fulfill these conditions for over 7.5M devices.  On the contrary,
applying those constraints to the set of devices that report at least one
malicious application, produces a larger population of users. This indeed
confirms the goodness of the model results and the importance of these two risk
indicators. Although less pronounced, we also observe the signers number to have
a much higher impact when looking at the clean subset: in this case, selecting devices
with more than 75 signers brings the percentage of considered devices down to
14.92\% while it keeps the one of the counterpart to 37.96\%.  For our initial task,
since applying all the constraints would result in an empty dataset, we decide to
relax the strictest ones (e.g., number of signers, alternative markets, and
percentage of apps downloaded from them): our group of clean devices with a high-security
profile accounts for 19k records (0.25\% of the total). 

\begin{table}[!h]
\small
	\caption{Number and fraction of clean and infected devices that fall in the riskiest bin of each feature.}
	\label{sec:systematic-tbl:cleanvsdirty}
	\setlength{\tabcolsep}{1.8pt}
\begin{tabular}{lrr}
\toprule

		\multirow{2}{*}{\textbf{Feature}}&
		\multicolumn{2}{c}{\textbf{Riskiest bins}} \\

		&
		\multicolumn{1}{c}{\textbf{Clean}}&
		\multicolumn{1}{c}{\textbf{Infected}}\\

\midrule
Applications                           &  1,680,125 (22.27 \%) &  467,676 (41.55 \%) \\
Activity days                          &  2,092,531 (27.73 \%) &  336,172 (29.87 \%) \\
Signers                       		   &  1,125,362 (14.92 \%) &  427,298 (37.96 \%) \\
Application categories                 &  1,986,501 (26.33 \%) &  188,374 (16.73 \%) \\
Update rate                            &  1,868,476 (24.76 \%) &  365,541 (32.47 \%) \\
\% Apps from AM       				   &     18,984 (0.25 \%)  &   36,809 (3.27 \%) \\
Alternative markets                    &        214 (0.01 \%)  &    1,671 (0.15 \%) \\
App prevalence                         &  1,771,623 (23.48 \%) &  395,999 (35.18 \%) \\
\bottomrule
\end{tabular}
\end{table}

On the contrary, to isolate \textit{at-risk} users in the former group, we first
check the percentage of malicious APKs with respect to the total number of
installed APKs on devices that report at least one malicious sample. Among
those, we then compute outliers by selecting users whose ratios exceed a
threshold identified by the 95\% quantile: out of 1.1M devices with at least one
detection, we select 63K devices whose malicious APKs represent at least 5\% of
the total number of applications. 

Overall, the risk-factor analysis of the best and worst devices confirms the higher relevance of a few indicators than others. In particular, the most marked difference is the one
related to the percentage of applications downloaded from alternative sources: while for
clean users this accounts for 8\%, for at-risk devices the percentage is above
42\%. In addition, we find the average number of application signers being $42$
for the best and $73$ for the worst devices, with the average for the whole population being around $50$.

\begin{figure}[!h]
\centering
  \includegraphics[width=\columnwidth]{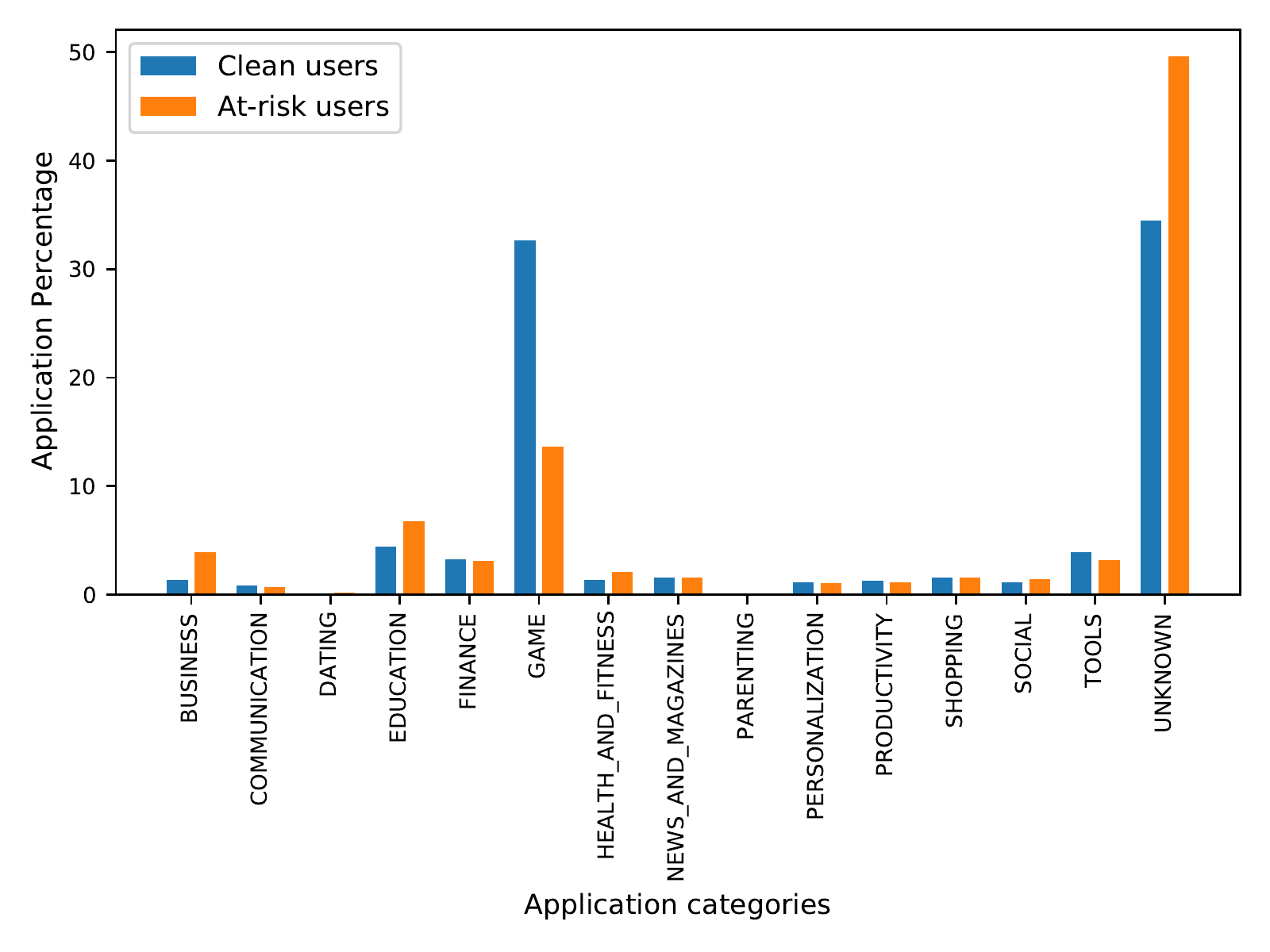}\hfill
	\caption{At-risk users, Clean and Average users categories distribution}
  \label{sec:appendix-fig:atRiskvsclean}
\end{figure}

To dive deeper into the differences between the two groups, we compute the
percentage of installed-application categories and plot their distribution in
Figure~\ref{sec:appendix-fig:atRiskvsclean}. For this task, we
do not consider in the computation the categories of malicious applications. The
analysis of the figure reveals a non-uniform distribution in the two
populations: remarkable cases are the high prevalence of \texttt{Unknown} apps
for at-risk users and the one of \texttt{Game} APKs into the group of clean
users.

\clearpage

\subsection{Dataset breakdown for prediction models}
\begin{table}[!h]
	\small
	\setlength{\tabcolsep}{6pt}
	\caption{Dataset size for each of the models used in our classification experiments.}
	\label{tbl:datasetSize}
	\begin{tabularx}{\columnwidth}{lrrr}
\toprule
		\multirow{2}{*}{\textbf{Model}}&
		\multicolumn{1}{c}{\textbf{PUA and}}&
		\multicolumn{1}{c}{\textbf{Malware}}&
		\multicolumn{1}{c}{\textbf{PUA}}\\

		&
		\multicolumn{1}{c}{\textbf{malware}}&
		\multicolumn{1}{c}{\textbf{only}}&
		\multicolumn{1}{c}{\textbf{only}}\\

\midrule
\rowcolor{red!15}Average users             	&      5,000 &          5,000 &     5,000 \\
\rowcolor{red!15}Mixed                     	&      5,000 &          5,000 &     5,000 \\
Game                      					&      5,000 &          5,000 &     5,000 \\
Video Players             					&      5,000 &          5,000 &     5,000 \\
Social                    					&      5,000 &          5,000 &     5,000 \\
Education                 					&      5,000 &          5,000 &     5,000 \\
Sports                    					&      5,000 &          1,476 &     5,000 \\
Finance                   					&      5,000 &          3,792 &     5,000 \\
News                      					&      5,000 &          1,646 &     5,000 \\
Business                  					&      5,000 &          2,928 &     5,000 \\
Health                    					&      5,000 &          1,458 &     5,000 \\
Weather                   					&      5,000 &          2,498 &     5,000 \\
Personalization           					&      5,000 &          3,764 &     5,000 \\
Communication             					&      5,000 &          3,084 &     5,000 \\
Navigation                					&      5,000 &          1,414 &     5,000 \\
Medical                   					&      5,000 &          1,702 &     5,000 \\
Books                     					&      5,000 &          1,924 &     5,000 \\
Lifestyle                 					&      3,496 &           518 &      3,184 \\
Photography               					&      5,000 &          2,658 &     5,000 \\
Shopping                  					&      5,000 &          1,870 &     5,000 \\
Music                     					&      5,000 &          1,728 &     5,000 \\
Auto                      					&      5,000 &          1,550 &     5,000 \\
Travel                    					&      5,000 &          1,042 &     5,000 \\
Entertainment             					&      5,000 &          1,198 &     5,000 \\
Comics                    					&      3,486 &           538 &      3,174 \\
Food                      					&      4,634 &           816 &      4,086 \\
Tools                     					&      5,000 &          2,240 &     5,000 \\
Communication-Tools       					&      5,000 &          1,110 &     4,506 \\
Productivity              					&      3,238 &           602 &      2,788 \\
House                     					&      2,638 &           388 &      2,396 \\
Social-Video Players      					&      5,000 &          1,980 &     5,000 \\
Entertainment-Game        					&      5,000 &          1,762 &     5,000 \\
Business-Finance          					&      5,000 &           980 &      4,540 \\
Education-Game            					&      5,000 &          1,246 &     5,000 \\
Comics-Game               					&      5,000 &           968 &      5,000 \\
Books-Education           					&      3,950 &           804 &      3,530 \\
Photography-Video Players 					&      5,000 &          1,112 &     4,970 \\
\bottomrule
\end{tabularx}
\end{table}

\newpage

\subsection{VirusTotal and Play Market coverage}
\begin{figure}[!h]
\centering
\includegraphics[width=0.9\columnwidth]{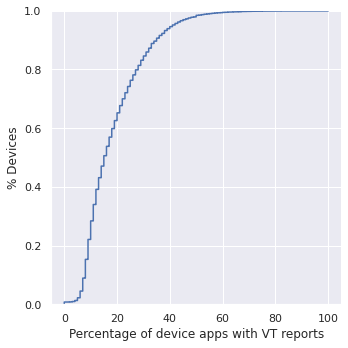}
\caption{Distribution of apps with VT reports per device in the filtered dataset.}
\label{sec:appendix-fig:vt_dev_distr}
\end{figure}

\begin{figure}[!h]
\centering
\includegraphics[width=0.9\columnwidth]{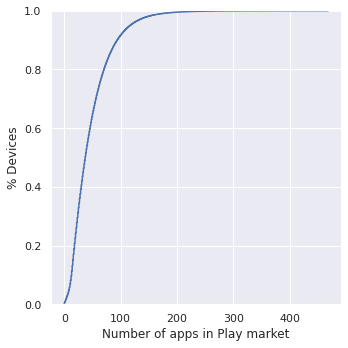}
\caption{Distribution of the number of apps in Google Play per device.}
\label{sec:appendix-fig:gplay_dev_distr}
\end{figure}

\subsection{Extended measurement results}
\begin{table*}[!h]
\centering
\footnotesize
\caption{Percentage distribution of devices capturing the number of application
	installed for each category. Values express percentages. Red cells
	represent the final selection. Values in blue represent the original
	selection based on the threshold that has been adapted to make it
	meaningful.}
	\label{tbl:profilingPercentage}
\begin{tabular}{lrrrrrrrrr}
\toprule
		\multicolumn{1}{c}{\textbf{Installed}}&
		\multicolumn{1}{c}{\textbf{Art and}}&
		\multicolumn{1}{c}{\textbf{Auto and}}&
		\multirow{2}{*}{\textbf{Beauty}} &  
		\multicolumn{1}{c}{\textbf{Books and}}&
		\multirow{2}{*}{\textbf{Business}} &  
		\multirow{2}{*}{\textbf{Comics}} &  
		\multirow{2}{*}{\textbf{Communication}} &  
		\multirow{2}{*}{\textbf{Dating}} &  
		\multirow{2}{*}{\textbf{Education}}\\

		\multicolumn{1}{c}{\textbf{Apps}}&
		\multicolumn{1}{c}{\textbf{design}}&
		\multicolumn{1}{c}{\textbf{vehicles}}&
		&  
		\multicolumn{1}{c}{\textbf{reference}}&
		&  
		&  
		&  
		&  
		\\
\midrule
1  & \B{7.24} 	& \B{5.71} 	&    \B{2.3} 	& 50.89 		&     46.11 	&   12.12 &          95.37 		&    \B{2.65} &      38.37 \\
2  & \R{1.03} 	& \R{4.34} 	&    \R{0.31} 	& 24.76 		&     24.66 	&    \R{1.93} &      83.9  		&    \R{0.75} &       20.3 \\
3  &  0.3 		& 1.51 		&    0.07 		& 12.03 		&     14.55 	&    0.91 &          69.39 		&    0.32 &      12.22 \\
4  & 0.12 		& 0.62 		&    0.02 		&  \R{5.56} 	&      \R{9.14} &    0.53 &          54.14 		&    0.16 &       \R{7.96} \\
5  & 0.06 		& 0.29 		&    0.01 		&   3.0 		&      5.95 	&    0.34 &          39.57 		&    0.09 &       5.47 \\
6  & 0.03 		& 0.15 		&     0.0 		&  1.77 		&      3.97 	&    0.23 &          28.1  		&    0.06 &       3.92 \\
7  & 0.02 		& 0.08 		&     0.0 		&  1.11 		&      2.66 	&    0.16 &          19.96 		&    0.04 &        2.9 \\
8  & 0.01 		& 0.05 		&     0.0 		&  0.74 		&      1.82 	&    0.12 &          13.88 		&    0.03 &       2.19 \\
9  & 0.01 		& 0.03 		&     0.0 		&  0.51 		&      1.26 	&    0.08 &           \R{9.37}	&    0.02 &        1.7 \\
10 &  0.0 		& 0.02 		&     0.0 		&  0.36 		&      0.88 	&    0.06 &           6.14 		&    0.01 &       1.34 \\
11 &  0.0 		& 0.01 		&     0.0 		&  0.27 		&      0.62 	&    0.05 &           3.92 		&    0.01 &       1.07 \\
12 &  0.0 		& 0.01 		&     0.0 		&   0.2 		&      0.45 	&    0.03 &           2.46 		&    0.01 &       0.87 \\
13 &  0.0 		& 0.01 		&     0.0 		&  0.16 		&      0.33 	&    0.03 &           1.54 		&    0.01 &       0.71 \\
14 &  0.0 		& 0.01 		&     0.0 		&  0.13 		&      0.24 	&    0.02 &           0.97 		&    0.01 &       0.59 \\
15 &  0.0 		&  0.0 		&     0.0 		&   0.1 		&      0.18 	&    0.02 &           0.62 		&     0.0 &        0.5 \\
16 &  0.0 		&  0.0 		&     0.0 		&  0.08 		&      0.14 	&    0.01 &           0.41 		&     0.0 &       0.42 \\
17 &  0.0 		&  0.0 		&     0.0 		&  0.07 		&       0.1 	&    0.01 &           0.27 		&     0.0 &       0.36 \\
18 &  0.0 		&  0.0 		&     0.0 		&  0.06 		&      0.08 	&    0.01 &           0.18 		&     0.0 &       0.31 \\
19 &  0.0 		&  0.0 		&     0.0 		&  0.05 		&      0.06 	&    0.01 &           0.13 		&     0.0 &       0.27 \\
20 &  0.0 		&  0.0 		&     0.0 		&  0.04 		&      0.05 	&     0.0 &           0.09 		&     0.0 &       0.23 \\
\bottomrule
\end{tabular}
\begin{tabular}{lrrrrrrrrr}
\toprule
		\multicolumn{1}{c}{\textbf{Installed}}&
		\multirow{2}{*}{\textbf{Entertainment}}&
		\multirow{2}{*}{\textbf{Events}}&
		\multirow{2}{*}{\textbf{Finance}} &  
		\multicolumn{1}{c}{\textbf{Food and}}&
		\multirow{2}{*}{\textbf{Game}} &  
		\multicolumn{1}{c}{\textbf{Health and}}&
		\multicolumn{1}{c}{\textbf{House and}}&
		\multicolumn{1}{c}{\textbf{Libraries and}}&
		\multirow{2}{*}{\textbf{Lifestyle}}\\

		\multicolumn{1}{c}{\textbf{Apps}}&
		&
		&
		&  
		\multicolumn{1}{c}{\textbf{drink}}&
		&  
		\multicolumn{1}{c}{\textbf{fitness}}&
		\multicolumn{1}{c}{\textbf{home}}&
		\multicolumn{1}{c}{\textbf{demo}}&
		\\
\midrule
1  &          81.72 		&    \B{3.62} 	&    73.29 &           39.96 		& 65.84 &               49.75 	&            \B{9.73} 	&                \B{8.41} 	&      63.51 \\
2  &          62.34 		&    \R{0.62} 	&    54.48 &           20.66 		& 51.31 &               24.48 	&            \R{2.45} 	&                 \R{1.0} 	&      39.33 \\
3  &          46.29 		&    0.16 		&    39.73 &           11.56 		& 41.52 &                13.4 		&            0.84 		&                0.66 		&      25.37 \\
4  &          33.74 		&    0.05 		&    28.83 &       \R{6.61} 	& 34.29 &                \R{7.98} 	&            0.33 		&                0.09 		&      16.87 \\
5  &          24.33 		&    0.01 		&    21.45 &            3.93 		& 28.84 &                5.01 		&            0.14 		&                0.03 		&      11.47 \\
6  &          17.46 		&     0.0 		&    16.17 &            2.43 &  24.6 &                3.28 		&            0.07 		& 0.02 		&       \R{7.91}\\
7  &          12.55 		&     0.0 		&    12.33 &            1.56 		& 21.23 &                2.22 		&            0.03 		&                 0.0 		&       5.51 \\
8  &           \R{9.01} 	&     0.0 		&     \R{9.48} &        1.02 		& 18.51 &                1.54 		&            0.02 		&                 0.0 		&       3.89 \\
9  &           6.49 		&     0.0 		&     7.36 &            0.69 		& 16.27 &                 1.1 		&            0.01 		&                 0.0 		&       2.78 \\
10 &           4.69 		&     0.0 		&     5.75 &            0.48 		& 14.41 &                 0.8 		&            0.01 		&                 0.0 		&       2.02 \\
11 &           3.41 		&     0.0 		&     4.53 &            0.34 		& 12.84 &                0.59 		&             0.0 		&                 0.0 		&       1.48 \\
12 &            2.5 		&     0.0 		&     3.59 &            0.25 		&  11.5 &                0.44 		&             0.0 		&                 0.0 		&        1.1 \\
13 &           1.85 		&     0.0 		&     2.87 &            0.18 		& 10.36 &                0.34 		&             0.0 		&                 0.0 		&       0.83 \\
14 &           1.38 		&     0.0 		&     2.31 &            0.14 		&  \R{9.37} &            0.26 		&             0.0 		&                 0.0 		&       0.64 \\
15 &           1.04 		&     0.0 		&     1.87 &             0.1 		&  8.51 &                 0.2 		&             0.0 		&                 0.0 		&       0.49 \\
16 &           0.79 		&     0.0 		&     1.53 &            0.08 		&  7.77 &                0.16 		&             0.0 		&                 0.0 		&       0.38 \\
17 &           0.61 		&     0.0 		&     1.25 &            0.06 		&  7.12 &                0.13 		&             0.0 		&                 0.0 		&        0.3 \\
18 &           0.48 		&     0.0 		&     1.04 &            0.05 		&  6.54 &                0.11 		&             0.0 		&                 0.0 		&       0.24 \\
19 &           0.37 		&     0.0 		&     0.86 &            0.04 		&  6.03 &                0.09 		&             0.0 		&                 0.0 		&       0.19 \\
20 &            0.3 		&     0.0 		&     0.72 &            0.03 		&  5.58 &                0.07 		&             0.0 		&                 0.0 		&       0.15 \\
\bottomrule
\end{tabular}
\end{table*}

\begin{table*}[!h]
\centering
\footnotesize
\begin{tabular}{lrrrrrrrrr}
\toprule
		\multicolumn{1}{c}{\textbf{Installed}}&
		\multicolumn{1}{c}{\textbf{Maps and}}&
		\multirow{2}{*}{\textbf{Medical}}&
		\multicolumn{1}{c}{\textbf{Music and}}&
		\multicolumn{1}{c}{\textbf{News and}}&
		\multirow{2}{*}{\textbf{Parenting}} &  
		\multirow{2}{*}{\textbf{Personalization}} &  
		\multirow{2}{*}{\textbf{Photography}} &  
		\multirow{2}{*}{\textbf{Productivity}} &  
		\multirow{2}{*}{\textbf{Shopping}} \\

		\multicolumn{1}{c}{\textbf{Apps}}&
		\multicolumn{1}{c}{\textbf{navigation}}&
		&
		\multicolumn{1}{c}{\textbf{audio}}&
		\multicolumn{1}{c}{\textbf{magazines}}&
		&  
		&  
		&  
		&  
		\\
\midrule
1  &                54.53 		&    17.42 &             77.9 &               65.11 &       \B{2.51} &            35.94 &        65.66 &         88.61 &     75.49 \\
2  &                28.98 		&     \R{5.58} &        53.35 &               37.65 &       \R{0.42} &            11.75 &        35.08 &         77.19 &     57.86 \\
3  &                14.13 		&     2.29 &            34.15 &                20.7 &       0.15 &             \R{5.96} &        18.68 &         66.01 &     43.86 \\
4  &                 \R{6.98} 	&     1.11 &             21.5 &               11.35 &       0.07 &             3.08 &        11.34 &         53.94 &     30.83 \\
5  &                 3.61 		&     0.61 &            13.51 &           \R{6.44} &    0.04 &             1.79 &         \R{7.05} &         43.82 &     22.29 \\
6  &                 1.96 		&     0.37 &             \R{8.51} &            3.82 &       0.02 &             1.18 &         4.64 &         34.78 &     16.32 \\
7  &                 1.12 		&     0.24 &             5.39 &                2.38 &       0.01 &             0.84 &         3.15 &          27.5 &     12.06 \\
8  &                 0.66 		&     0.17 &             3.44 &                1.54 &       0.01 &             0.63 &         2.18 &         21.74 &      \R{8.98} \\
9  &                 0.41 		&     0.12 &             2.24 &                1.04 &        0.0 &              0.5 &         1.55 &         17.03 &      6.74 \\
10 &                 0.26 		&      0.1 &             1.48 &                0.72 &        0.0 &              0.4 &         1.13 &         13.17 &       5.1 \\
11 &                 0.17 		&     0.08 &              1.0 &                0.51 &        0.0 &             0.33 &         0.85 &         10.07 &      3.89 \\
12 &                 0.11 		&     0.06 &             0.69 &                0.37 &        0.0 &             0.27 &         0.65 &          \R{7.68} &      2.98 \\
13 &                 0.08 		&     0.05 &             0.49 &                0.28 &        0.0 &             0.23 &          0.5 &          5.85 &       2.3 \\
14 &                 0.05 		&     0.04 &             0.35 &                0.21 &        0.0 &              0.2 &         0.39 &          4.46 &      1.78 \\
15 &                 0.04 		&     0.03 &             0.26 &                0.16 &        0.0 &             0.17 &         0.31 &          3.42 &      1.38 \\
16 &                 0.03 		&     0.03 &              0.2 &                0.12 &        0.0 &             0.15 &         0.25 &          2.63 &      1.08 \\
17 &                 0.02 		&     0.02 &             0.15 &                 0.1 &        0.0 &             0.13 &         0.21 &          2.03 &      0.85 \\
18 &                 0.02 		&     0.02 &             0.12 &                0.08 &        0.0 &             0.12 &         0.17 &          1.57 &      0.67 \\
19 &                 0.01 		&     0.02 &              0.1 &                0.06 &        0.0 &             0.11 &         0.14 &          1.22 &      0.54 \\
20 &                 0.01 		&     0.01 &             0.08 &                0.05 &        0.0 &              0.1 &         0.12 &          0.96 &      0.43 \\
\bottomrule
\end{tabular}
\begin{tabular}{lrrrrrrrrr}
\toprule
		\multicolumn{1}{c}{\textbf{Installed}}&
		\multirow{2}{*}{\textbf{Social}}&
		\multirow{2}{*}{\textbf{Sports}}&
		\multirow{2}{*}{\textbf{Tools}}&
		\multicolumn{1}{c}{\textbf{Travel and}}&
		\multicolumn{1}{c}{\textbf{Video}}&
		\multirow{2}{*}{\textbf{Weather}}\\

		\multicolumn{1}{c}{\textbf{Apps}}&
		&
		&
		&
		\multicolumn{1}{c}{\textbf{local}}&
		\multicolumn{1}{c}{\textbf{players}}&
		\\
\midrule
1  &   77.98 		&   25.36 		&  94.48 &             67.12 &          66.59 &    34.96 \\
2  &   53.81 		&   10.62 		&  85.36 &             43.57 &          38.88 &    11.75 \\
	3  &   31.16 		&    \R{5.77} 	&  74.66 &             29.22 &          19.46 &     \R{4.59} \\
4  &   17.41 		&    3.49 		&  64.33 &             20.33 &           \R{9.68} &     2.07 \\
5  &    \R{9.38} 	&    2.24 		&  55.14 &             14.58 &           5.03 &     1.05 \\
6  &     5.0 		&    1.48 		&  46.92 &             10.67 &           2.71 &     0.58 \\
7  &    2.71 		&    1.01 		&  39.66 &              \R{7.95} &       1.52 &     0.35 \\
8  &    1.52 		&     0.7 		&  33.46 &              6.01 &            0.9 &     0.22 \\
9  &    0.89 		&     0.5 		&  28.13 &               4.6 &           0.56 &     0.14 \\
10 &    0.55 		&    0.36 		&  23.62 &              3.55 &           0.36 &      0.1 \\
11 &    0.36 		&    0.26 		&  19.88 &              2.78 &           0.24 &     0.07 \\
12 &    0.24 		&    0.19 		&  16.81 &              2.19 &           0.16 &     0.05 \\
13 &    0.16 		&    0.14 		&  14.25 &              1.73 &           0.12 &     0.04 \\
14 &    0.12 		&    0.11 		&  12.11 &              1.38 &           0.08 &     0.03 \\
15 &    0.09 		&    0.08 		&   10.3 &              1.11 &           0.06 &     0.02 \\
16 &    0.06 		&    0.06 		&   \R{8.72} &          0.89 &           0.05 &     0.02 \\
17 &    0.05 		&    0.05 		&   7.32 &              0.73 &           0.04 &     0.02 \\
18 &    0.04 		&    0.04 		&    6.1 &              0.59 &           0.03 &     0.01 \\
19 &    0.03 		&    0.03 		&   5.07 &              0.49 &           0.02 &     0.01 \\
20 &    0.02 		&    0.02 		&   4.19 &               0.4 &           0.02 &     0.01 \\
\bottomrule
\end{tabular}
\end{table*}

\begin{table*}[!t]
	\small
	\centering
	\caption{Summary of malicious application encounters for user profiles with at least 10k devices}
	\label{tbl:profileMalicious_10K}
	\begin{tabular}{lr|rrr|rr}
\toprule
		\multirow{2}{*}{\textbf{Profile}}&
		\multirow{2}{*}{\textbf{Size}}&
		\multicolumn{1}{c}{\textbf{Encounter}}&
		\multicolumn{1}{c}{\textbf{Malware}}&
		\multicolumn{1}{c}{\textbf{PUA}}&
		\multicolumn{1}{c}{\textbf{Malicious APKs}}&
		\multicolumn{1}{c}{\textbf{Malicious Encounters}}\\
		&
		&
		\multicolumn{1}{c}{\textbf{malicious apps}}&
		\multicolumn{1}{c}{\textbf{only}}&
		\multicolumn{1}{c}{\textbf{only}}&
		\multicolumn{1}{c}{\textbf{from profile apps}}&
		\multicolumn{1}{c}{\textbf{from profile apps}}\\
\midrule
\rowcolor{red!15}Average users             &  599,483 &    51,885 (8.65\%) &        12.01\% &   73.14\% &            	   &        		   \\
\rowcolor{red!15}Mixed                     &  470,383 &    93,407 (19.86\%) &       10.92\% &   79.70\% &            	   &                   \\
Game                      &  235,434 &    48,903 (20.77\%) &       10.53\% &   79.22\% &           34.97\% &   21,100 (43.15\%) \\
Video Players             &  106,723 &    27,543 (25.81\%) &        8.73\% &   79.29\% &           17.25\% &    8,245 (29.94\%) \\
Social                    &   99,176 &    18,497 (18.65\%) &       10.96\% &   81.68\% &            1.62\% &      434 (2.35\%) \\
Education                 &   97,405 &    14,551 (14.94\%) &       11.11\% &   76.41\% &            2.26\% &      710 (4.88\%) \\
Sports                    &   83,891 &     6,253 (7.45\%) &         9.58\% &   85.06\% &            2.05\% &      213 (3.41\%) \\
Finance                   &   80,617 &    11,457 (14.21\%) &       13.42\% &   79.18\% &            8.35\% &      934 (8.15\%) \\
News                      &   77,068 &     6,595 (8.56\%) &         9.01\% &   82.58\% &            1.77\% &      173 (2.62\%) \\
Business                  &   74,982 &     8,561 (11.42\%) &       10.55\% &   79.30\% &            2.86\% &      410 (4.79\%) \\
Health                    &   68,111 &     5,440 (7.99\%) &        10.51\% &   83.42\% &           13.97\% &      810 (14.89\%) \\
Weather                   &   66,864 &     7,930 (11.86\%) &        8.35\% &   82.17\% &           49.84\% &    3,622 (45.67\%) \\
Personalization           &   62,944 &     9,076 (14.42\%) &       13.14\% &   75.31\% &            1.62\% &      272 (3.00\%) \\
Communication             &   60,478 &     8,939 (14.78\%) &       12.84\% &   78.67\% &            4.93\% &      498 (5.57\%) \\
Navigation                &   59,425 &     5,048 (8.49\%) &        11.49\% &   82.19\% &            1.14\% &       67 (1.33\%) \\
Medical                   &   57,605 &     5,128 (8.90\%) &        11.88\% &   80.05\% &            0.46\% &       41 (0.80\%) \\
Books                     &   57,493 &     5,865 (10.20\%) &       12.86\% &   79.54\% &            5.61\% &      404 (6.89\%) \\
Lifestyle                 &   56,966 &     2,320 (4.07\%) &         8.66\% &   85.69\% &           12.37\% &      301 (12.97\%) \\
Photography               &   53,813 &     8,439 (15.68\%) &       10.50\% &   81.17\% &           16.47\% &    1,498 (17.75\%) \\
Shopping                  &   49,444 &     5,824 (11.78\%) &       13.72\% &   80.72\% &            5.30\% &      296 (5.08\%) \\
Music                     &   47,992 &     6,837 (14.25\%) &        9.38\% &   84.29\% &           23.18\% &    1,523 (22.28\%) \\
Auto                      &   42,114 &     4,736 (11.25\%) &       11.57\% &   80.43\% &            1.35\% &       90 (1.90\%) \\
Travel                    &   41,834 &     4,087 (9.77\%) &        10.06\% &   83.85\% &            7.05\% &      498 (12.18\%) \\
Entertainment             &   40,848 &     5,023 (12.30\%) &        9.91\% &   84.49\% &            6.67\% &      393 (7.82\%) \\
Comics                    &   36,419 &     1,970 (5.41\%) &         9.59\% &   83.35\% &            0.87\% &       21 (1.07\%) \\
Food                      &   33,164 &     2,747 (8.28\%) &        11.90\% &   82.27\% &            0.78\% &       37 (1.35\%) \\
Tools                     &   32,549 &     4,778 (14.68\%) &       15.93\% &   71.16\% &           24.64\% &    1,670 (34.95\%) \\
Communication-Tools       &   23,632 &     3,154 (13.35\%) &       13.67\% &   77.65\% &           23.21\% &      914 (28.98\%) \\
Productivity              &   22,056 &     2,020 (9.16\%) &        13.86\% &   80.79\% &            2.52\% &       63 (3.12\%) \\
House                     &   19,935 &     1,634 (8.20\%) &        10.04\% &   84.09\% &            2.82\% &       58 (3.55\%) \\
Social-Video Players      &   17,184 &     6,034 (35.11\%) &        7.41\% &   81.27\% &           23.42\% &    2,248 (37.26\%) \\
Entertainment-Game        &   16,998 &     5,354 (31.50\%) &        9.30\% &   77.94\% &           46.12\% &    2,909 (54.33\%) \\
Business-Finance          &   16,242 &     3,000 (18.47\%) &       14.10\% &   78.60\% &           12.88\% &      410 (13.67\%) \\
Education-Game            &   14,769 &     3,633 (24.60\%) &        9.33\% &   77.92\% &           27.01\% &    1,542 (42.44\%) \\
Comics-Game               &   13,549 &     3,317 (24.48\%) &        8.38\% &   81.49\% &           40.62\% &    1,646 (49.62\%) \\
Books-Education           &   12,024 &     2,267 (18.85\%) &       12.31\% &   77.06\% &            6.63\% &      297 (13.10\%) \\
Photography-Video Players &   10,993 &     3,123 (28.41\%) &       10.79\% &   78.67\% &           24.32\% &      955 (30.58\%) \\
\midrule
Overall average 		  &    6,846 &     	 990 (14.55\%) &       10.67\% &   79.20\% &           13.72\% &     151 (16.55\%) \\
\bottomrule
\end{tabular}
\end{table*}

\begin{table*}
	\small
	\centering
	\caption{Top-4 categories responsible for malware encounters for user profiles with at least 10k devices.}
	\label{tbl:threatSources_10K}
\begin{tabular}{lrrrr}
\toprule

		\multirow{2}{*}{\textbf{Profile}}&
		\multicolumn{4}{c}{\textbf{Top-categories}}\\
		&
		\multicolumn{1}{c}{\textbf{1st}}&
		\multicolumn{1}{c}{\textbf{2nd}}&
		\multicolumn{1}{c}{\textbf{3rd}}&
		\multicolumn{1}{c}{\textbf{4th}}\\
\midrule
\rowcolor{red!15}Average users             &          Tools (30.35\%) &           Game (16.83\%) &  Video Players (16.23\%) &  Entertainment (5.90\%) \\
\rowcolor{red!15}Mixed                     &          Tools (25.37\%) &           Game (13.51\%) &  Video Players (10.00\%) &          Music (8.35\%) \\
Game                      &           Game (64.04\%) &           Tools (8.02\%) &   Entertainment (7.56\%) &  Video Players (5.42\%) \\
Video Players             &   Video Players (46.9\%) &          Tools (19.27\%) &            Game (7.53\%) &          Music (4.44\%) \\
Social                    &           Tools (31.0\%) &  Video Players (22.05\%) &            Game (9.83\%) &          Music (5.06\%) \\
Education                 &          Tools (24.54\%) &           Game (15.49\%) &  Video Players (13.33\%) &      Education (8.12\%) \\
Sports                    &          Tools (23.57\%) &           Game (11.97\%) &           Music (8.49\%) &    Photography (8.34\%) \\
Finance                   &          Tools (28.33\%) &        Finance (15.25\%) &  Video Players (10.11\%) &           Game (8.67\%) \\
News                      &          Tools (28.92\%) &        Weather (28.52\%) &  Video Players (10.56\%) &           Game (7.86\%) \\
Business                  &           Tools (26.2\%) &  Video Players (13.27\%) &           Game (10.54\%) &       Business (9.86\%) \\
Health                    &         Health (30.21\%) &          Tools (20.16\%) &     Photography (7.63\%) &           Game (7.57\%) \\
Weather                   &        Weather (66.84\%) &           Tools (9.75\%) &     Photography (4.23\%) &  Entertainment (3.17\%) \\
Personalization           &          Tools (28.77\%) &           Game (15.57\%) &          Music (11.56\%) &  Video Players (9.19\%) \\
Communication             &          Tools (25.62\%) &           Game (19.93\%) &  Communication (10.95\%) &  Video Players (6.44\%) \\
Navigation                &          Tools (28.88\%) &         Weather (12.4\%) &            Game (8.23\%) &    Photography (7.58\%) \\
Medical                   &          Tools (22.89\%) &           Game (10.38\%) &   Video Players (8.19\%) &    Photography (8.16\%) \\
Books                     &          Tools (22.05\%) &          Books (13.46\%) &           Game (13.23\%) &          Music (7.58\%) \\
Lifestyle                 &          Tools (23.06\%) &      Lifestyle (15.06\%) &           Game (11.29\%) &       Weather (11.18\%) \\
Photography               &    Photography (33.93\%) &          Tools (21.31\%) &   Video Players (7.84\%) &  Entertainment (7.83\%) \\
Shopping                  &          Tools (22.79\%) &        Shopping (11.3\%) &             Game (8.8\%) &         Health (8.63\%) \\
Music                     &          Music (44.47\%) &          Tools (18.46\%) &            Game (7.09\%) &  Entertainment (5.12\%) \\
Auto                      &          Tools (26.39\%) &           Game (11.92\%) &           Music (9.88\%) &    Photography (8.35\%) \\
Travel                    &           Tools (26.4\%) &         Travel (16.72\%) &     Photography (9.72\%) &          Music (6.34\%) \\
Entertainment             &           Tools (23.5\%) &           Game (15.04\%) &  Entertainment (14.19\%) &  Video Players (9.19\%) \\
Comics                    &           Game (31.64\%) &          Tools (17.38\%) &  Video Players (15.66\%) &          Music (6.97\%) \\
Food                      &          Tools (26.57\%) &           Game (13.43\%) &   Video Players (6.97\%) &          Music (6.23\%) \\
Tools                     &          Tools (54.64\%) &            Game (7.38\%) &     Photography (4.59\%) &          Music (4.57\%) \\
Communication-Tools       &          Tools (45.99\%) &            Game (8.85\%) &           Music (5.86\%) &    Photography (5.71\%) \\
Productivity              &          Tools (28.13\%) &     Photography (10.8\%) &            Game (8.12\%) &         Health (7.88\%) \\
House                     &          Tools (21.64\%) &    Photography (12.39\%) &            Game (9.51\%) &          Music (8.81\%) \\
Social-Video Players      &  Video Players (50.46\%) &           Tools (21.8\%) &            Game (5.68\%) &  Entertainment (3.49\%) \\
Entertainment-Game        &           Game (55.54\%) &  Entertainment (17.99\%) &           Tools (7.98\%) &          Music (5.18\%) \\
Business-Finance          &          Tools (31.08\%) &         Finance (13.2\%) &       Business (11.58\%) &  Video Players (8.72\%) \\
Education-Game            &           Game (56.05\%) &          Tools (10.62\%) &   Entertainment (6.03\%) &  Video Players (5.08\%) \\
Comics-Game               &           Game (67.02\%) &   Video Players (8.61\%) &           Tools (7.51\%) &          Music (5.12\%) \\
Books-Education           &          Tools (23.64\%) &            Game (11.2\%) &      Education (10.47\%) &  Video Players (9.38\%) \\
Photography-Video Players &  Video Players (26.33\%) &    Photography (24.07\%) &          Tools (17.84\%) &          Music (6.74\%) \\
\midrule
Overall average           &          Tools (22.19\%) &           Game (12.58\%) &           Music (7.27\%) &  Video Players (7.18\%) \\
\bottomrule
\end{tabular}
\end{table*}

\begin{table*}[!t]
    \centering
    \small
    \caption{Comparison of the odds ratios of installing PUA and/or malware
	between the baseline and the last bin for the average users, mixed profile,
	and the profiles with at least 10k users.}
	\label{sec:behavioral2:profileodds}
	\setlength{\tabcolsep}{3pt}

    \begin{tabular}{lrrrrrrr}
    \toprule
		\multirow{2}{*}{\textbf{Feature}}&
		\multicolumn{1}{c}{\textbf{Average}}&
		\multirow{2}{*}{\textbf{Mixed}}&
		\multirow{2}{*}{\textbf{Game}}&
		\multirow{2}{*}{\textbf{Video Players}}&
		\multirow{2}{*}{\textbf{Social}}&
		\multirow{2}{*}{\textbf{Education}}&
		\multirow{2}{*}{\textbf{Sports}}\\

		&
		\multicolumn{1}{c}{\textbf{users}}&
		&
		&
		&
		&
		&
		\\
\midrule
Applications       	&  1.39 &   \underline{0.69} 	&  1.48 & 1.19 & 1.27 				&  1.17 			& 1.22 \\
Activity days       &  1.50 &   1.32 				&  1.25 & 1.66 & 1.46 				&  1.30 			& 1.19 \\
Signers             &  3.64 &   3.26 				&  3.05 & 2.53 & 2.38 				&  3.88 			& 5.19 \\
Categories          &  0.41 &   0.47 				&  0.70 & 0.42 & 0.52 				&  0.67 			& 0.41 \\
Update Rate         &  3.28 &   1.70 				&  2.33 & 2.21 & 1.48 				&  1.78 			& 1.54 \\
\% apps from AM 	&  4.14 &  \textbf{26.78} 		&  6.53 & 2.10 & 1.88 				& \textbf{50.87} 	& 5.46 \\
Alternative Markets	&  1.28 &   2.36 				&  2.63 & 2.44 & 1.65 				&  2.23 			& \underline{0.71} \\
Avg Prevalence      &  0.61 &   0.70 				&  0.75 & 0.87 & 0.48 				&  0.68 			& 0.81 \\
Countries           &  1.26 &   1.18 				&  1.09 & 1.55 & \underline{0.30} 	&  0.64 			& 1.01 \\
\bottomrule
	\end{tabular}
    \begin{tabular}{lrrrrrrr}
    \toprule
		\multirow{2}{*}{\textbf{Feature}}&
		\multirow{2}{*}{\textbf{Finance}}&
		\multicolumn{1}{c}{\textbf{News and}}&
		\multirow{2}{*}{\textbf{Business}}&
		\multicolumn{1}{c}{\textbf{Health and}}&
		\multirow{2}{*}{\textbf{Weather}}&
		\multirow{2}{*}{\textbf{Personalization}}&
		\multirow{2}{*}{\textbf{Communication}}\\

		&
		&
		\multicolumn{1}{c}{\textbf{Magazines}}&
		&
		\multicolumn{1}{c}{\textbf{Fitness}}&
		&
		&
		\\
\midrule
Applications        & 1.16 &  1.20 				&   \underline{0.49} 	&   1.21 			&  1.28 & \underline{0.71} 	& \underline{0.29} \\
Activity days       & 1.16 &  1.45 				&   1.28 				&   1.15 			&  1.11 & 1.11 				& \underline{0.90} \\
Signers             & 3.27 &  3.23 				&   5.28 				&   3.47 			&  2.23 & 6.15 				& 3.63 \\
Categories          & 0.72 &  0.51 				&   \underline{1.28} 	&   0.52 			&  0.55 & 0.39 				& 0.61 \\
Update Rate         & 1.13 &  2.42 				&   1.36 				&   1.41 			&  2.25 & 1.93 				& 1.48 \\
\% apps from AM 	& 2.38 &  5.02 				&  	\textbf{27.69} 		&  \textbf{71.42} 	&  6.44 & 7.92 				& 3.01 \\
Alternative Markets & 1.24 &  \underline{0.92} 	&   1.31 				&   1.35 			&  1.05 & 3.62 				& 1.55 \\
Avg Prevalence      & 0.65 &  0.71 				&   0.66 				&   0.71 			&  0.83 & 0.57 				& 0.61 \\
Countries           & 1.37 &  1.31 				&   1.15 				&   1.01 			&  1.10 & 1.19 				& 1.38 \\
\bottomrule
\end{tabular}
    \begin{tabular}{lrrrrrrr}
    \toprule
		\multirow{2}{*}{\textbf{Feature}}&
		\multicolumn{1}{c}{\textbf{Maps and}}&
		\multirow{2}{*}{\textbf{Medical}}&
		\multicolumn{1}{c}{\textbf{Books and}}&
		\multirow{2}{*}{\textbf{Lifestyle}}&
		\multirow{2}{*}{\textbf{Photography}}&
		\multirow{2}{*}{\textbf{Shopping}}&
		\multicolumn{1}{c}{\textbf{Music and}}\\

		&
		\multicolumn{1}{c}{\textbf{Navigation}}&
		&
		\multicolumn{1}{c}{\textbf{Reference}}&
		&
		&
		&
		\multicolumn{1}{c}{\textbf{Audio}}\\
\midrule
Applications        & 1.30 &  1.35 				&   1.47 				&   \underline{0.86} 	&  \underline{0.53} & 1.00 &   1.14 \\
Activity days       & 1.03 &  1.03 				&   1.01 				&   \underline{0.95} 	&  1.03 			& 1.13 &   1.19 \\
Signers             & 4.77 &  4.16 				&   4.58 				&   5.20 				&  7.51 			& 4.42 &   3.51 \\
Categories          & 0.43 &  0.61 				&   0.35 				&   0.57 				&  0.29 			& 0.48 &   0.38 \\
Update Rate         & 1.39 &  1.42 				&   1.77 				&   1.10 				&  1.52 			& 1.23 &   1.58 \\
\% apps from AM 	& 8.13 &  \textbf{25.21} 	&   9.28 				&   \textbf{147.32} 	&  5.16 			& 2.04 &   9.09 \\
Alternative Markets & 1.12 &  \underline{0.86} 	&   \underline{0.74} 	&   2.14 				&  2.32 			& 1.19 &   1.04 \\
Avg Prevalence      & 0.84 &  0.69 				&   0.64 				&   0.65 				&  0.87 			& 0.80 &   0.68 \\
Countries           & 1.04 &  1.18 				&   1.15 				&   1.19 				&  1.14 			& 1.12 &   1.07 \\
\bottomrule
\end{tabular}
    \begin{tabular}{lrrrrrrr}
    \toprule
		\multirow{2}{*}{\textbf{Feature}}&
		\multicolumn{1}{c}{\textbf{Auto and}}&
		\multicolumn{1}{c}{\textbf{Travel and}}&
		\multirow{2}{*}{\textbf{Entertainment}}&
		\multirow{2}{*}{\textbf{Comics}}&
		\multicolumn{1}{c}{\textbf{Food and}}&
		\multirow{2}{*}{\textbf{Tools}}&
		\multicolumn{1}{c}{\textbf{Communication}}\\

		&
		\multicolumn{1}{c}{\textbf{Vehicles}}&
		\multicolumn{1}{c}{\textbf{Locals}}&
		&
		&
		\multicolumn{1}{c}{\textbf{Drinks}}&
		&
		\multicolumn{1}{c}{\textbf{Tools}}\\
\midrule
Applications        &  1.37 			&    1.56 			& \underline{0.95} 	&    \underline{0.63} 	&  1.04 			&   \underline{0.14} 	& \underline{0.36} \\
Activity days       &  1.24 			&    1.03 			& 1.65 				&    1.83 				&  \underline{0.96} &   1.02 				& 1.03 \\
Signers             &  3.30 			&    3.21 			& 5.12 				&    7.10 				&  6.84 			&   4.24 				& 4.64 \\
Categories          &  0.57 			&    0.64 			& 0.49 				&    0.21 				&  0.40 			&   0.58 				& 0.57 \\
Update Rate         &  1.50 			&    1.01 			& 2.20 				&    4.57 				&  1.31 			&   1.94 				& 1.77 \\
\% apps from AM 	& \textbf{24.74} 	&   \textbf{40.49} 	& 1.38 				&    7.28 				& \textbf{22.88} 	&   3.62 				& 1.78 \\
Alternative Markets &  \underline{0.93} &    1.69 			& \underline{0.89} 	&    \underline{0.32} 	&  1.19 			&   2.72 				& 5.94 \\
Avg Prevalence      &  0.72 			&    0.81 			& 0.77 				&    0.76 				&  0.85 			&   0.53 				& 0.74 \\
Countries           &  1.00 			&    1.18 			& 1.16 				&    1.34 				&  1.02 			&   1.28 				& 1.46 \\
\bottomrule
\end{tabular}
\end{table*}
\clearpage
\begin{table*}[!t]
    \centering
    \small
	\setlength{\tabcolsep}{3pt}

    \begin{tabular}{lrrrrrrr}
    \toprule
		\multirow{2}{*}{\textbf{Feature}}&
		\multirow{2}{*}{\textbf{Productivity}}&
		\multicolumn{1}{c}{\textbf{House and}}&
		\multicolumn{1}{c}{\textbf{Social and}}&
		\multicolumn{1}{c}{\textbf{Entertainment}}&
		\multicolumn{1}{c}{\textbf{Business}}&
		\multicolumn{1}{c}{\textbf{Education}}&
		\multicolumn{1}{c}{\textbf{Comics}}\\

		&
		&
		\multicolumn{1}{c}{\textbf{Home}}&
		\multicolumn{1}{c}{\textbf{Video Players}}&
		\multicolumn{1}{c}{\textbf{Game}}&
		\multicolumn{1}{c}{\textbf{Finance}}&
		\multicolumn{1}{c}{\textbf{Game}}&
		\multicolumn{1}{c}{\textbf{Game}}\\
\midrule
Applications        &          \underline{0.52} &   1.30 			& 1.55 & 1.02 & \underline{0.86} 	&  1.11 			&         1.06 \\
Activity days       &          1.49 			&   1.11 			& 1.88 & 1.33 &  1.13 				&  1.03 			&         1.31 \\
Signers             &          4.21 			&   4.22 			& 1.42 & 3.91 &  3.28 				&  4.15 			&         3.47 \\
Categories          &          0.56 			&   0.45 			& 0.79 & 0.65 &  0.85 				&  0.78 			&         0.56 \\
Update Rate         &          1.56 			&   1.45 			& 1.15 & 1.87 &  1.05 				&  1.79 			&         3.23 \\
\% apps from AM 	&         11.56 			&  \textbf{46.29} 	& 1.71 & 2.82 & 11.38 				& \textbf{33.08} 	&         3.41 \\
Alternative Markets &          \underline{0.73} &   1.29 			& 2.94 & 2.62 &  1.78 				&  3.46 			&         3.63 \\
Avg Prevalence      &          0.62 			&   0.61 			& 0.63 & 0.77 &  0.73 				&  0.77 			&         0.80 \\
Countries           &          1.03 			&   1.10 			& 0.40 & 1.17 &  1.41 				&  1.08 			&         1.03 \\
\bottomrule
	\end{tabular}
    \begin{tabular}{lrrrrrrr}
    \toprule
		\multirow{2}{*}{\textbf{Feature}}&
		\multicolumn{1}{c}{\textbf{Books}}&
		\multicolumn{1}{c}{\textbf{Photography and}}\\

		&
		\multicolumn{1}{c}{\textbf{Education}}&
		\multicolumn{1}{c}{\textbf{Video Players}}\\
\midrule
Applications        &  1.68 			&  \underline{0.71} \\
Activity days       &  1.04 			&  \underline{0.98} \\
Signers             &  2.53 			&  3.60 \\
Categories          &  0.67 			&  0.41 \\
Update Rate         &  1.43 			&  1.48 \\
\% apps from AM 	&  \textbf{19.48} 	&  3.72 \\
Alternative Markets &  1.65 			&  4.41 \\
Avg Prevalence      &  0.69 			&  0.85 \\
Countries           &  1.15 			&  1.15 \\
\bottomrule
\end{tabular}
\end{table*}

\begin{table*}
	\centering
	\caption{Classification accuracy of whole-population and per-profile models
	tested on average users, mixed profile, and on the user profiles with at
	least 10k users.}
        \label{tbl:classification_appendix}
\small
\begin{tabular}{l|rrr|rrr|r}
\toprule
		\multirow{3}{*}{\textbf{Profile}}&
		\multicolumn{3}{c|}{\textbf{Whole-population model}}&
		\multicolumn{3}{c|}{\textbf{Per-profile models}}&
		\multirow{3}{*}{\textbf{Avg improvement}}\\
		&
		\multicolumn{1}{c}{\textbf{PUA or}}&
		\multicolumn{1}{c}{\textbf{Malware}}&
		\multicolumn{1}{c|}{\textbf{PUA}}&
		\multicolumn{1}{c}{\textbf{PUA}}&
		\multicolumn{1}{c}{\textbf{Malware}}&
		\multicolumn{1}{c|}{\textbf{PUA}}&
		\\
		&
		\multicolumn{1}{c}{\textbf{Malware}}&
		\multicolumn{1}{c}{\textbf{only}}&
		\multicolumn{1}{c|}{\textbf{only}}&
		\multicolumn{1}{c}{\textbf{Malware}}&
		\multicolumn{1}{c}{\textbf{only}}&
		\multicolumn{1}{c|}{\textbf{only}}&
		\\
\midrule
\rowcolor{red!15}Average users             &                   51.90\% &                       49.50\% &                   51.50\% &                      84.35\% &                          83.56\% &                      77.91\% &            30.97\% \\
\rowcolor{red!15}Mixed                     &                   50.30\% &                       51.40\% &                   52.90\% &                      72.16\% &                          68.42\% &                      67.03\% &            17.67\% \\
Game                      &                   53.50\% &                       49.60\% &                   48.80\% &                      64.38\% &                          64.22\% &                      72.22\% &            16.31\% \\
Video Players             &                   48.50\% &                       50.10\% &                   53.10\% &                      73.20\% &                          82.16\% &                      67.24\% &            23.63\% \\
Social                    &                   52.30\% &                       49.60\% &                   47.30\% &                      54.76\% &                          75.29\% &                      70.21\% &            17.02\% \\
Education                 &                   50.90\% &                       49.90\% &                   49.70\% &                      71.33\% &                          82.81\% &                      78.68\% &            27.44\% \\
Sports                    &                   48.30\% &                       63.51\% &                   48.30\% &                      71.05\% &                          75.00\% &                      71.05\% &            19.00\% \\
Finance                   &                   49.40\% &                       61.00\% &                   50.40\% &                      56.90\% &                          72.46\% &                      56.82\% &             8.46\% \\
News                      &                   52.20\% &                       58.79\% &                   48.50\% &                      76.81\% &                          75.00\% &                      75.00\% &            22.44\% \\
Business                  &                   51.50\% &                       68.26\% &                   52.10\% &                      78.91\% &                          77.85\% &                      82.11\% &            22.34\% \\
Health                    &                   51.40\% &                       64.38\% &                   48.80\% &                      69.07\% &                          71.15\% &                      69.39\% &            15.01\% \\
Weather                   &                   51.50\% &                       66.40\% &                   49.60\% &                      63.64\% &                          64.58\% &                      75.53\% &            12.08\% \\
Personalization           &                   51.00\% &                       67.33\% &                   49.60\% &                      69.31\% &                          69.46\% &                      77.78\% &            16.21\% \\
Communication             &                   52.10\% &                       61.59\% &                   48.20\% &                      59.68\% &                          75.00\% &                      73.08\% &            15.29\% \\
Navigation                &                   51.10\% &                       57.95\% &                   47.90\% &                      67.14\% &                          64.86\% &                      64.41\% &            13.15\% \\
Medical                   &                   48.20\% &                       63.05\% &                   49.60\% &                      70.21\% &                          83.33\% &                      66.30\% &            19.66\% \\
Books                     &                   46.10\% &                       70.39\% &                   49.70\% &                      72.22\% &                          71.05\% &                      76.06\% &            17.71\% \\
Lifestyle                 &                   63.86\% &                       63.46\% &                   65.46\% &                      70.59\% &                          55.00\% &                      68.33\% &             0.38\% \\
Photography               &                   48.40\% &                       68.05\% &                   51.20\% &                      68.83\% &                          68.13\% &                      75.81\% &            15.04\% \\
Shopping                  &                   49.60\% &                       60.96\% &                   49.40\% &                      67.31\% &                          57.58\% &                      66.00\% &            10.31\% \\
Music                     &                   51.10\% &                       57.81\% &                   52.30\% &                      70.79\% &                          58.00\% &                      68.06\% &            11.88\% \\
Auto                      &                   51.40\% &                       61.29\% &                   50.70\% &                      66.28\% &                          77.27\% &                      73.68\% &            17.95\% \\
Travel                    &                   51.90\% &                       66.03\% &                   51.20\% &                      71.91\% &                          58.62\% &                      76.83\% &            12.74\% \\
Entertainment             &                   49.50\% &                       70.00\% &                   48.90\% &                      64.29\% &                          71.43\% &                      72.06\% &            13.13\% \\
Comics                    &                   67.34\% &                       70.37\% &                   67.56\% &                      70.00\% &                          81.25\% &                      75.00\% &             6.99\% \\
Food                      &                   61.60\% &                       66.46\% &                   60.64\% &                      74.65\% &                          65.62\% &                      75.00\% &             8.86\% \\
Tools                     &                   47.50\% &                       64.96\% &                   51.00\% &                      71.43\% &                          67.71\% &                      74.31\% &            16.66\% \\
Communication-Tools       &                   51.20\% &                       69.82\% &                   64.86\% &                      74.03\% &                          59.38\% &                      66.67\% &             4.73\% \\
Productivity              &                   62.81\% &                       66.94\% &                   62.01\% &                      62.79\% &                          58.33\% &                      72.22\% &             0.53\% \\
House                     &                   55.87\% &                       64.11\% &                   51.66\% &                      66.67\% &                          75.00\% &                      70.00\% &            13.34\% \\
Social-Video Players      &                   50.40\% &                       64.64\% &                   48.30\% &                      82.43\% &                          78.35\% &                      69.23\% &            22.22\% \\
Entertainment-Game        &                   50.10\% &                       62.89\% &                   49.80\% &                      68.00\% &                          63.16\% &                      61.29\% &             9.89\% \\
Business-Finance          &                   49.60\% &                       58.67\% &                   62.33\% &                      62.12\% &                          53.85\% &                      70.59\% &             5.32\% \\
Education-Game            &                   45.80\% &                       70.00\% &                   51.10\% &                      74.04\% &                          62.90\% &                      69.47\% &            13.17\% \\
Comics-Game               &                   48.00\% &                       70.11\% &                   49.70\% &                      75.76\% &                          77.27\% &                      81.25\% &            22.16\% \\
Books-Education           &                   61.78\% &                       63.98\% &                   61.61\% &                      80.95\% &                          70.00\% &                      81.58\% &            15.05\% \\
Photography-Video Players &                   53.00\% &                       63.23\% &                   61.27\% &                      66.27\% &                          80.00\% &                      62.90\% &            10.56\% \\
\midrule
Overall average           &                   52.19\% &                       62.34\% &                   52.89\% &                      69.84\% &                          70.14\% &                      71.65\% &            14.74\% \\
\bottomrule
\end{tabular}
\end{table*}

\begin{table*}[t]
	\centering
	\setlength{\tabcolsep}{3pt}
	\footnotesize
	\caption{Top-5 features of the classifiers trained on the whole-population, 
	average users, mixed profile, and on the user profiles with at least 10k users.}
	\label{sec:prediction-tbl:features:appendix}
	\begin{tabular}{crrrrrrr}
\toprule
		\textbf{\Centerstack{Feature\\rank}}&
		\textbf{\Centerstack{Unique\\model}}&
		\textbf{\Centerstack{Average users}}&
		\textbf{\Centerstack{Mixed}}&
		\textbf{\Centerstack{Game}}&
		\textbf{\Centerstack{Video\\Players}}&
		\textbf{\Centerstack{Social}}&
		\textbf{\Centerstack{Education}}\\
\midrule
1st &      Signers - 9.76 &     Unknown - 19.93 &    Avg prev. - 7.56 &      Unknown - 16.5 &     Unknown - 10.83 &     Unknown - 10.83 &  Avg prev. - 11.32 \\
2nd &      Unknown - 8.93 &    Avg prev. - 9.92 &       Signers - 6.9 &          Apps - 8.6 &      Signers - 9.02 &    Avg prev. - 8.65 &     Unknown - 9.96 \\
3rd &   \% apps AM - 5.61 &      Signers - 7.11 &  Update rate - 6.19 &       Signers - 6.7 &    Avg prev. - 7.32 &         Days - 6.48 &   \% apps AM - 9.19 \\
4th &         Apps - 5.39 &  Update rate - 6.91 &         Days - 5.89 &    Avg prev. - 6.33 &         Apps - 6.97 &         Apps - 5.64 &     Signers - 8.99 \\
5th &  Update rate - 5.15 &         Apps - 5.66 &         Apps - 4.67 &  Update rate - 5.77 &  Update rate - 5.78 &      Signers - 5.19 &        Days - 6.77 \\
\bottomrule
\end{tabular}
\begin{tabular}{crrrrrrr}
\toprule
		\textbf{\Centerstack{Feature\\rank}}&
		\textbf{\Centerstack{Sports}}&
		\textbf{\Centerstack{Finance}}&
		\textbf{\Centerstack{News and\\magazines}}&
		\textbf{\Centerstack{Business}}&
		\textbf{\Centerstack{Health and\\fitness}}&
		\textbf{\Centerstack{Weather}}&
		\textbf{\Centerstack{Personalization}}\\
\midrule
1st &      Unknown - 8.14 &  Update rate - 7.71 &      Unknown - 8.92 &   Avg prev. - 10.41 &  Update rate - 8.25 &         Days - 6.88 &    Avg prev. - 9.93 \\
2nd &  Update rate - 6.98 &          Days - 6.7 &         Days - 6.64 &      Signers - 9.49 &    Avg prev. - 8.21 &  Update rate - 6.56 &      Unknown - 9.24 \\
3rd &         Days - 6.81 &      Unknown - 6.66 &         Apps - 6.49 &    \% apps AM - 8.35 &      Unknown - 6.44 &      Unknown - 6.23 &      Signers - 8.76 \\
4th &      Signers - 6.47 &    Avg prev. - 6.08 &  Update rate - 6.09 &         Days - 7.56 &         Days - 6.41 &    Avg prev. - 6.23 &    \% apps AM - 6.87 \\
5th &    Avg prev. - 5.95 &         Apps - 5.28 &      Signers - 6.04 &  Update rate - 6.29 &      Signers - 6.09 &         Apps - 5.67 &         Apps - 5.36 \\
\bottomrule
\end{tabular}
\begin{tabular}{crrrrrrr}
\toprule
		\textbf{\Centerstack{Feature\\rank}}&
		\textbf{\Centerstack{Communication}}&
		\textbf{\Centerstack{Maps and\\navigation}}&
		\textbf{\Centerstack{Medical}}&
		\textbf{\Centerstack{Books and\\reference}}&
		\textbf{\Centerstack{Lifestyle}}&
		\textbf{\Centerstack{Photography}}&
		\textbf{\Centerstack{Shopping}}\\
\midrule
1st &      Signers - 8.54 &         Days - 6.62 &      Unknown - 8.66 &    Avg prev. - 8.07 &  Update rate - 8.37 &      Unknown - 8.72 &  Update rate - 7.9 \\
2nd &      Unknown - 7.61 &      Unknown - 6.54 &    Avg prev. - 7.82 &      Unknown - 8.01 &      Signers - 7.44 &    Avg prev. - 7.85 &        Days - 6.31 \\
3rd &    Avg prev. - 7.37 &      Signers - 6.53 &      Signers - 7.53 &  Update rate - 7.66 &         Apps - 6.37 &      Signers - 7.33 &   Avg prev. - 5.88 \\
4th &         Apps - 6.49 &    Avg prev. - 6.31 &         Apps - 6.31 &      Signers - 7.42 &    Avg prev. - 6.34 &  Update rate - 5.87 &        Apps - 5.19 \\
5th &  Update rate - 5.36 &  Update rate - 5.86 &  Update rate - 6.29 &         Apps - 5.85 &      Unknown - 6.29 &         Apps - 5.41 &     Unknown - 4.74 \\
\bottomrule
\end{tabular}
\begin{tabular}{crrrrrrr}
\toprule
		\textbf{\Centerstack{Feature\\rank}}&
		\textbf{\Centerstack{Music and\\audio}}&
		\textbf{\Centerstack{Auto and\\vehicles}}&
		\textbf{\Centerstack{Travel and\\local}}&
		\textbf{\Centerstack{Entertainment}}&
		\textbf{\Centerstack{Comics}}&
		\textbf{\Centerstack{Food and\\drink}}&
		\textbf{\Centerstack{Tools}}\\
\midrule
1st &     Avg prev. - 7.5 &      Unknown - 7.51 &     Signers - 7.56 &          Days - 7.3 &      Unknown - 9.69 &   Avg prev. - 7.77 &      Signers - 9.37 \\
2nd &  Update rate - 6.65 &    Avg prev. - 7.34 &   Avg prev. - 7.25 &  Update rate - 6.07 &  Update rate - 8.18 &     Unknown - 7.77 &      Unknown - 8.29 \\
3rd &         Days - 6.58 &  Update rate - 7.08 &   \% apps AM - 6.23 &      Signers - 5.95 &      Signers - 6.72 &     Signers - 7.42 &    Avg prev. - 8.23 \\
4th &      Unknown - 5.36 &         Days - 6.81 &     Unknown - 6.07 &       Unknown - 5.8 &    Avg prev. - 5.87 &        Days - 6.77 &         Apps - 7.23 \\
5th &      Signers - 5.27 &      Signers - 5.88 &        Apps - 5.58 &    Avg prev. - 5.64 &         Apps - 5.63 &  Update rate - 6.7 &  Update rate - 5.21 \\
\bottomrule
\end{tabular}
\begin{tabular}{crrrrrrr}
\toprule
		\textbf{\Centerstack{Feature\\rank}}&
		\textbf{\Centerstack{Communication\\Tools}}&
		\textbf{\Centerstack{Productivity}}&
		\textbf{\Centerstack{House and\\home}}&
		\textbf{\Centerstack{Social\\Video Players}}&
		\textbf{\Centerstack{Entertainment\\Game}}&
		\textbf{\Centerstack{Business\\Finance}}&
		\textbf{\Centerstack{Education\\Game}}\\
\midrule
1st &      Signers - 9.67 &         Days - 8.92 &   Avg prev. - 9.93 &      Unknown - 11.69 &     Unknown - 8.82 &  Update rate - 7.73 &      Unknown - 8.87 \\
2nd &    Avg prev. - 7.17 &    Avg prev. - 7.61 &     Signers - 7.51 &    Avg prev. - 11.16 &     Signers - 8.57 &         Days - 6.97 &      Signers - 7.43 \\
3rd &      Unknown - 6.08 &  Update rate - 6.94 &     Unknown - 6.57 &          Apps - 6.76 &         Apps - 7.3 &    Avg prev. - 6.78 &  Update rate - 7.21 \\
4th &         Apps - 5.32 &      Signers - 6.31 &  Update rate - 5.9 &       Signers - 6.22 &        Days - 7.07 &      Unknown - 5.23 &         Apps - 6.68 \\
5th &        Tools - 4.74 &        Tools - 4.18 &        Apps - 5.49 &   Update rate - 4.64 &        Game - 6.22 &         Apps - 4.82 &         Days - 6.09 \\
\bottomrule
\end{tabular}
\begin{tabular}{crrrr}
\toprule
		\textbf{\Centerstack{Feature\\rank}}&
		\textbf{\Centerstack{Comics\\Game}}&
		\textbf{\Centerstack{Books and reference\\Education}}&
		\textbf{\Centerstack{Photography\\Video Players}}&
		\textbf{\Centerstack{Overall\\average}}\\
\midrule
1st &     Unknown - 12.78 &   Avg prev. - 10.27 &          Avg prev. - 8.52 &      Unknown - 8.18 \\
2nd &      Signers - 7.07 &      Unknown - 9.85 &            Signers - 7.48 &    Avg prev. - 7.51 \\
3rd &         Apps - 6.45 &         Days - 7.54 &            Unknown - 6.99 &       Signers - 7.1 \\
4th &     Avg prev. - 5.9 &       Signers - 7.0 &               Apps - 5.52 &  Update rate - 5.86 \\
5th &         Game - 5.85 &    \% apps AM - 6.97 &        Update rate - 5.48 &         Apps - 5.39 \\
\bottomrule
\end{tabular}
\end{table*}

\end{document}